\documentclass[12pt,twoside, a4paper]{article}
\pdfoutput=1
\def\pd{\partial}
\def\mc{\mathcal}

\def\ul{\underline}
\def\bb{\mathbb}

\usepackage[dvips]{graphicx}
\usepackage{amssymb}
\usepackage{amssymb,amsmath}
\usepackage{graphicx}
\usepackage{dsfont}
\usepackage{caption}
\usepackage{subcaption}
\usepackage{mathtools}
\usepackage{verbatim}
\usepackage{graphicx}
\usepackage{multirow}
\usepackage[outline]{contour}
\usepackage{xcolor,colortbl}
\usepackage{pdflscape}
\input{epsf.sty}  
\begin{document}
\begin{center}
\LARGE{\textbf{Janus solutions from dyonic $ISO(7)$ maximal gauged supergravity}}
\end{center}
\vspace{1 cm}
\begin{center}
\large{\textbf{Parinya Karndumri}$^a$ and \textbf{Chawakorn Maneerat}$^b$}
\end{center}
\begin{center}
String Theory and Supergravity Group, Department
of Physics, Faculty of Science, Chulalongkorn University, 254 Phayathai Road, Pathumwan, Bangkok 10330, Thailand
\end{center}
E-mail: $^a$parinya.ka@hotmail.com \\
E-mail: $^b$chawakorn.manee@gmail.com
\vspace{1 cm}\\
\begin{abstract}
We give a large class of new supersymmetric Janus solutions in dyonic $ISO(7)$ maximal gauged supergravity encompassing all known solutions appearing recently. We consider $SO(3)$ invariant sector in which the gauged supergravity admits four supersymmetric $AdS_4$ vacua with $N=1,1,2,3$ unbroken supersymmetries and $G_2$, $SU(3)$, $SU(3)\times U(1)$ and $SO(4)$ symmetries, respectively. We find Janus solutions preserving $N=1,2,3$ supersymmetries and interpolating between $AdS_4$ vacua and the $N=8$ SYM phase. These solutions can be interpreted as conformal interfaces between SYM and various conformal phases of the dual field theory in three dimensions. The solutions can also be embedded in massive type IIA theory by a consistent truncation on $S^6$. We also find a number of singular Janus solutions interpolating between $N=8$ SYM or $AdS_4$ vacua and singularities or between singular geometries. It turns out that, apart from the SYM phase on D2-branes, all the singularities arising in these solutions are unphysical in both four- and ten-dimensional frameworks. 
\end{abstract}
\newpage
\section{Introduction}
Janus solutions in gauged supergravity holographically describe conformal interfaces or defects within the dual conformal field theories (CFTs) via the AdS/CFT correspondence \cite{maldacena,Gubser_AdS_CFT,Witten_AdS_CFT}. These solutions take the form of $AdS_{D-1}$-sliced domain walls which are asymptotic to $AdS_D$ geometries on both sides. Due to the presence of the $AdS_{D-1}$ slice in the $D$-dimensional bulk, the conformal symmetry of CFT$_{D-1}$ dual to the $AdS_D$ fixed point is broken to a smaller conformal symmetry of CFT$_{D-2}$ on the defect dual to the $AdS_{D-1}$ slice. The first solution of this type has been found in \cite{Bak_Janus} by considering a deformation of the $AdS_5\times S^5$ geometry in type IIB theory. The solution describes a conformal interface in the dual $N=4$ Super Yang-Mills (SYM) theory. Other solutions have been extensively studied in both type IIB theory and five-dimensional gauged supergravity, see for example \cite{Freedman_Janus,DHoker_Janus,Witten_Janus,Freedman_Holographic_dCFT,5D_Janus_CK,5D_Janus_DHoker1,
5D_Janus_DHoker2,5D_Janus_Suh,Bobev_5D_Janus1,Bobev_5D_Janus2}. Similar solutions in other dimensions have also been given in \cite{warner_Janus,N3_Janus,tri-sasakian-flow,orbifold_flow,Minwoo_4DN8_Janus,Kim_Janus,N5_flow,N6_flow,3D_Janus_de_Boer,3D_Janus_Bachas,3D_Janus_Bak,half_BPS_AdS3_S3_ICFT,exact_half_BPS_string,multi_face_Janus,
4D_Janus_from_11D,6D_Janus,3D_Janus,N8_omega_Janus,N4_Janus}. 
\\
\indent In this paper, we are interested in supersymmetric Janus solutions from dyonically gauged $N=8$ supergravity with $ISO(7)\sim CSO(7,0,1)\sim SO(7)\ltimes \mathbf{R}^7$ gauge group. This gauged supergravity has been constructed in \cite{ISO7_Guarino} by using the embedding tensor formalism. The ten-dimensional origin from massive type IIA theory and the corresponding dual three-dimensional SCFTs in terms of Chern-Simons-Matter (CSM) theory have also been given in \cite{mIIA_S6_Guariano1}. In addition, the full consistent truncation of massive type IIA theory on $S^6$ has been constructed in \cite{mIIA_S6_Guariano2}. Unlike the well-known $SO(8)$ $N=8$ gauged supergravity, the dyonic $ISO(7)$ gauged supergravity does not admit a maximally supersymmetric $AdS_4$ vacuum but a half-supersymmetric domain wall dual to an effective $N=8$ Super Yang-Mills (SYM) theory on D2-branes. However, this gauged supergravity does admit supersymmetric $AdS_4$ vacua with $N=1,1,2,3$ supersymmetries and $G_2$, $SU(3)$, $U(3)$ and $SO(4)$ symmetries, respectively \cite{ISO7_Guarino,mIIA_S6_Guariano1,G2_AdS4,AdS4_Henning}, see also \cite{Romans_mass_flow} and \cite{Bobev_AdS4_ISO7} for $N=1$ $AdS_4$ vacua with $U(1)$ symmetry and without any residual symmetry. These vacua are dual to different conformal phases of the $N=8$ SYM in the presence of Chern-Simons terms. The corresponding holographic RG flows from the $N=8$ SYM to these critical points have already been studied recently in \cite{ISO7_N3_flow}, see also \cite{Romans_mass_flow} for an earlier result.
\\
\indent The study of Janus solutions in this gauged supergravity has been initiated in \cite{Minwoo_4DN8_Janus} in which a number of solutions within $SU(3)$ and $G_2$ truncations have been given. We will extend this study by including a large number of new Janus solutions within these two sectors and solutions from the $SO(3)$ and $SO(3)\times U(1)$ invariant sectors which have not been considered in \cite{Minwoo_4DN8_Janus}. In particular, we will find Janus solutions involving $N=3$ $AdS_4$ vacuum with $SO(4)$ symmetry and solutions preserving $N=2,3$ supersymmetries. These solutions are entirely new, and by including the $N=3$ critical point, we find a large number of supersymmetric Janus solutions interpolating among $AdS_4$ vacua and $N=8$ SYM phase. With the $S^6$ consistent truncation, the solutions obtained in this paper can be embedded in massive type IIA theory leading to complete holographic descriptions of conformal interfaces within the dual three-dimensional SYM and SCFTs.  
\\
\indent The paper is organized as follows. In section \ref{N8_SUGRA},
we review the structure of four-dimensional $N=8$ gauged supergravity with dyonic $ISO(7)$ gauge group and supersymmetric $AdS_4$ vacua with $N=1,1,2,3$ supersymmetries and $G_2$, $SU(3)$, $U(3)$ and $SO(4)$ symmetries within $SO(3)$ invariant sector. In sections \ref{N3Janus}, \ref{N2Janus} and \ref{N1Janus}, a large number of supersymmetric Janus solutions preserving $N=3,2,1$ supersymmetries are given. Conclusions and comments are given in section \ref{conclusion}.

\section{$N=8$ gauged supergravity with dyonic $ISO(7)$ gauge group}\label{N8_SUGRA}
We first give a brief review of four-dimensional $N=8$ gauged supergravity with dyonic $ISO(7)$ gauge group constructed in \cite{ISO7_Guarino} to which we refer for more detail. We will mostly follow the conventions of \cite{ISO7_Guarino}. There is only one supermultiplet in $N=8$ supersymmetry, the supergravity multiplet, with the field content
\begin{equation}
(e^{\hat{\mu}}_\mu,\psi^I_\mu,A^{AB}_\mu,\chi_{IJK},\Sigma_{IJKL}).
\end{equation}
This consists of the graviton $e^{\hat{\mu}}_\mu$, $8$ gravitini $\psi^I_\mu$, $28$ vectors $A^{AB}_\mu=-A^{BA}_\mu$, $56$ spin-$\frac{1}{2}$ fields $\chi_{IJK}=\chi_{[IJK]}$ and $70$ scalars $\Sigma_{IJKL}=\Sigma_{[IJKL]}$. Throughout the paper, space-time and tangent space indices are denoted by $\mu,\nu,\ldots =0,1,2,3$ and $\hat{\mu},\hat{\nu},\ldots =0,1,2,3$, respectively. 
\\
\indent The $N=8$ supergravity admits global $E_{7(7)}$ and local composite $SU(8)$ symmetries with the corresponding fundamental representations are respectively described by indices $\bb{M},\bb{N},\ldots=1,2,3,\ldots ,56$ and $I,J,K,\ldots =1,2,3,\ldots 8$. The scalars $\Sigma_{IJKL}$ parametrizing the $E_{7(7)}/SU(8)$ coset manifold can be described by the coset representative ${\mc{V}_{\bb{M}}}^{\ul{\bb{M}}}$. The local $SU(8)$ indices $\ul{\bb{M}},\ul{\bb{N}},\ldots$ will be decomposed as $_{\ul{\bb{M}}}=(_{[IJ]},^{[IJ]})$. Similarly, the global $E_{7(7)}$ indices $\bb{M},\bb{N},\ldots$ will be decomposed in the $SL(8)$ basis as $_{\bb{M}}=(_{[AB]},^{[AB]})$ with indices $A,B,\ldots =1,2,3,\ldots, 8$ denoting fundamental representation of $SL(8)\subset E_{7(7)}$. The scalars $\Sigma_{IJKL}$ are self-dual 
\begin{equation}
\Sigma_{IJKL}=\frac{1}{4!}\epsilon_{IJKLMNPQ}\Sigma^{MNPQ}
 \end{equation}
with $\Sigma^{IJKL}=(\Sigma_{IJKL})^*$ and $\epsilon_{IJKLMNPQ}$ being the invariant tensor of the $SU(8)$ R-symmetry. 
\\
\indent The split of an index $_{\ul{\bb{M}}}=(_{[IJ]},^{[IJ]})$ is in accord with the decomposition of $E_{7(7)}$ fundamental representation $\mathbf{56}$ under $SL(8)$ as $\mathbf{56}\rightarrow \mathbf{28}+\mathbf{28}'$. In $SL(8)$ basis, $E_{7(7)}$ generators are decomposed as 
\begin{equation}
t_\alpha ={t_A}^B\otimes t_{ABCD},\qquad \alpha=1,2,3,\ldots, 133
\end{equation}
with ${t_A}^B$ being $SL(8)$ generators. This realizes the decomposition of $E_{7(7)}$ adjoint representation $\mathbf{133}\rightarrow \mathbf{63}+\mathbf{70}$. An explicit representation of these generators in fundamental representation of $E_{7(7)}$ is given by
\begin{equation}
{({t_A}^B)_{[CD]}}^{[EF]}=4\delta^B_{[C}\delta^{EF}_{D]A}+\frac{1}{2}\delta^B_A\delta^{EF}_{CD}\quad \textrm{and}\quad {({t_A}^B)^{[EF]}}_{[CD]}=-{({t_A}^B)_{[CD]}}^{[EF]}
\end{equation}
for $SL(8)$ generators in $\mathbf{63}$ and 
\begin{equation}
(t_{ABCD})_{[EF][GH]}=\frac{2}{4!}\epsilon_{ABCDEFGH}\quad \textrm{and}\quad (t_{ABCD})^{[EF][GH]}=2\delta^{EFGH}_{ABCD}
\end{equation}
for the remaining generators in $\mathbf{70}$. $\delta^{AB\dots}_{CD\ldots}=\delta^{[AB\dots]}_{[CD\ldots]}$ is antizymmetrized with weight one.
\\
\indent The action of the global $E_{7(7)}$ symmetry includes electric-magnetic duality. The vector fields $A_\mu^{AB}$ together with the magnetic dual $A_{\mu AB}$ transform in the fundamental $\mathbf{56}$ representation. In general, the Lagrangian of the ungauged $N=8$ supergravity will exhibit only particular subgroups of $E_{7(7)}$ depending on the electric-magnetic or symplectic frames. The full $E_{7(7)}$ symmetry is realized only through the field equations and Bianchi identities. The most general gaugings of the $N=8$ supergravity can be described by the embedding tensor ${\Theta_{\bb{M}}}^\alpha$ which introduces a minimal coupling for various fields in the ungauged supergravity via the covariant derivative
\begin{equation}
D_\mu=\nabla_\mu-gA^{\bb{M}}_\mu {\Theta_{\bb{M}}}^\alpha t_\alpha
 \end{equation} 
with $\nabla_\mu$ being the usual space-time covariant derivative. The gauge generators $X_{\bb{M}}={\Theta_{\bb{M}}}^\alpha t_\alpha$ must form a closed subalgebra of $E_{7(7)}$ leading to the quadratic constraint of the form 
\begin{equation}
\Omega^{\bb{M}\bb{N}}{\Theta_{\bb{M}}}^\alpha{\Theta_{\bb{N}}}^\beta=0\, .
\end{equation}
$\Omega^{\bb{M}\bb{N}}$ is the symplectic form of the duality group $Sp(56,\bb{R})$ in which $E_{7(7)}$ is embedded. Moreover, supersymmetry requires the embedding tensor to transform as $\mathbf{912}$ representation of $E_{7(7)}$. 
\\
\indent The embedding tensor for $ISO(7)$ gauge group is given by
\begin{equation}
{{\Theta_{[AB]}}^C}_D=2\delta^C_{[A}\theta_{B]D}\qquad \textrm{and}\qquad {\Theta^{[AB]C}}_D=2\delta^{[A}_D\xi^{B]C}
 \end{equation}
 with  
\begin{equation}
\theta=\left(\begin{array}{cc}
\delta_{ij}&0\\
0&0
\end{array}\right)\qquad \textrm{and}\qquad \xi=\left(\begin{array}{cc}
0_{7\times 7}&0\\
0&c
\end{array}\right)
\end{equation}
for a constant $c$. It should also be remarked that any values of $c\neq0$ are equivalent \cite{omega_N8_2}. The $SL(8)$ fundamental indices are decomposed as $A,B,\ldots =(i,8)$. The full embedding tensor can be explicitly written as
\begin{eqnarray}
{{\Theta_{[ij]}}^k}_l&=&2\delta^k_{[i}\delta_{j]l},\qquad {{\Theta_{i8}}^8}_k=-\delta_{ik},\nonumber \\
{\Theta^{[ij]k}}_l&=&0,\qquad {\Theta^{[i8]8}}_k=c\delta^i_k\, . \label{embedding_tensor}
\end{eqnarray}
In this paper, we are only interested in supersymmetric Janus solutions involving only the metric and scalars. Therefore, we will set all the vector and tensor fields as well as the fermions to zero from now on.  
\\
\indent The gauged Lagrangian can be written as
\begin{equation}
e^{-1}\mc{L}=\frac{1}{2}R+\frac{1}{96}\textrm{Tr}(\pd_\mu \mc{M}\pd^\mu \mc{M}^{-1})-V
\end{equation}
with the scalar matrix $\mc{M}=\mc{V}\mc{V}^T$ and the scalar potential given by
\begin{equation}
V=\frac{1}{672}g^2({X_{\bb{M}\bb{N}}}^{\bb{R}}{X_{\bb{P}\bb{Q}}}^{\bb{S}}\mc{M}^{\bb{M}\bb{P}}\mc{M}^{\bb{N}\bb{Q}}\mc{M}_{\bb{R}\bb{S}}+7{X_{\bb{M}\bb{N}}}^{\bb{Q}}{X_{\bb{P}\bb{Q}}}^{\bb{N}}\mc{M}^{\bb{M}\bb{P}}).
\end{equation}
The tensor ${X_{\bb{M}\bb{N}}}^{\bb{P}}$ is defined by
\begin{equation}
{X_{\bb{M}\bb{N}}}^{\bb{P}}={\Theta_{\bb{M}}}^\alpha {(t_\alpha)_{\bb{N}}}^{\bb{P}}={{\Theta_{\bb{M}}}^C}_D{({t_C}^D)_{\bb{N}}}^{\bb{P}}\, .
\end{equation}
With the embedding tensor \eqref{embedding_tensor}, various components of ${X_{\bb{M}\bb{N}}}^{\bb{P}}$ are given by
\begin{eqnarray}
{X_{[ij][kl]}}^{[mn]}&=&-X_{[ij]\phantom{[mn]}[kl]}^{\phantom{[ij]}[mn]}=-8\delta^{[m}_{[i}\delta_{j][k}\delta^{n]}_{l]},\nonumber \\
{X_{[ij][k8]}}^{[m8]}&=&-X_{[ij]\phantom{[m8]}[k8]}^{\phantom{[ij]}[m8]}=-2\delta^{m}_{[i}\delta_{j]k},\nonumber \\
{X_{[k8][ij]}}^{[m8]}&=&-X_{[k8]\phantom{[m8]}[ij]}^{\phantom{[k8]}[m8]}=-2\delta_{k[i}\delta^{m}_{j]},\nonumber \\
X^{[k8]\phantom{[ij]}[m8]}_{\phantom{[k8]}[ij]}&=&-{X^{[k8][m8]}}_{[ij]}=2c\delta^{k}_{[i}\delta^{m}_{j]}\, .
\end{eqnarray}
\indent The supersymmetry transformations of $\psi^I_\mu$ and $\chi_{IJK}$ relevant for finding supersymmetric solutions are given by
\begin{eqnarray}
\delta\psi^I_\mu&=&2D_\mu \epsilon^I+\sqrt{2}gA_1^{IJ}\gamma_\mu \epsilon_J,\\
\delta \chi^{IJK}&=&-2\sqrt{2}P_\mu^{IJKL}\gamma^\mu \epsilon_L-2g{A_{2L}}^{IJK}\epsilon^L\, .
\end{eqnarray}
The covariant derivative of $\epsilon^I$ is defined by
\begin{equation}
D_\mu \epsilon^I=\pd_\mu \epsilon^I+\frac{1}{4}{\omega_\mu}^{\hat{\mu}\hat{\nu}}\gamma_{\hat{\mu}\hat{\nu}}\epsilon^I+\frac{1}{2}{{Q_\mu}^I}_J\epsilon^J\, .
\end{equation}
The composite connection ${{Q_\mu}^I}_J=({Q_{\mu I}}^J)^*$ and the vielbein $P_\mu^{IJKL}$ on the $E_{7(7)}/SU(8)$ coset are given by
\begin{eqnarray}
{Q_{\mu I}}^J&=&\frac{i}{3}(\mc{V}_{ABIK}\pd_\mu \mc{V}^{JK}-{\mc{V}^{AB}}_{IK}\pd_\mu {\mc{V}_{AB}}^{JK}),\\
P_{\mu IJKL}&=&\frac{i}{2}(\mc{V}_{ABIJ}\pd_\mu {\mc{V}^{AB}}_{KL}-{\mc{V}^{AB}}_{IJ}\pd_\mu \mc{V}_{ABKL})
\end{eqnarray}
with 
\begin{equation}
P_\mu^{IJKL}=\frac{1}{4!}\epsilon^{IJKLMNPQ}P_{\mu MNPQ}\, .
\end{equation}
$A_1$ and $A_2$ tensors are written in terms of the T-tensor defined by 
\begin{equation}
{T_{\ul{\bb{M}}\ul{\bb{N}}}}^{\ul{\bb{P}}}={\mc{V}_{\ul{\bb{M}}}}^{\bb{M}}{\mc{V}_{\ul{\bb{N}}}}^{\bb{N}}{\mc{V}_{\bb{P}}}^{\ul{\bb{P}}}{X_{\bb{M}\bb{N}}}^{\bb{P}}
\end{equation}
via the following relations
\begin{equation}
A_1^{IJ}=\frac{4}{21}{T^{IKJL}}_{KL}\qquad \textrm{and}\qquad {A_{2I}}^{JKL}=2{T_{MI}}^{MJKL}\, .
\end{equation}
In these equations, we have written ${\mc{V}_{\ul{\bb{M}}}}^{\bb{M}}={(\mc{V}^{-1})_{\ul{\bb{M}}}}^{\bb{M}}$.

\subsection{$SO(3)$ invariant truncation and supersymmetric $AdS_4$ vacua} 
The resulting $ISO(7)$ gauged supergavity with $c\neq 0$ admits a number of supersymmetric $AdS_4$ vacua. Four of these vacua can be found by considering $SO(3)$ truncation of the full $E_{7(7)}/SU(8)$ scalar manifold. This truncation leads to half-maximal gauged supergavity coupled to three vector multiplets with $ISO(3)\times SO(3)$ gauge group \cite{N4_from_ISO7}. There are eight singlet scalars corresponding to the following $E_{7(7)}$ non-compact generators
\begin{eqnarray}
Y_1&=&{t_4}^4+{t_6}^6-{t_2}^2-{t_8}^8,\qquad Y_2={t_2}^2-{t_4}^4+{t_6}^6-{t_8}^8,\nonumber \\
Y_3&=&{t_2}^2+{t_4}^4-{t_6}^6-{t_8}^8,\qquad Y_4={t_1}^1-{t_2}^2+{t_3}^3-{t_4}^4+{t_5}^5-{t_6}^6+{t_7}^7-{t_8}^8,\nonumber \\
\tilde{Y}_1&=&t_{1238}+t_{2578},\qquad \tilde{Y}_2=t_{1458}+t_{4738},\nonumber \\
\tilde{Y}_3&=&t_{1678}+t_{6358},\qquad \tilde{Y}_4=t_{8246}\, .
\end{eqnarray}
The coset representative is then given by
\begin{equation}
\mc{V}=e^{-12(2\zeta_1\tilde{Y}_1+2\zeta_2\tilde{Y}_2+2\zeta_3 \tilde{Y}_3+\chi \tilde{Y}_4)}e^{\frac{1}{4}(2\phi_1Y_1+2\phi_2Y_2+2\phi_3Y_3+\varphi Y_4)}\, .
\end{equation}
\indent At this point, we note the values of various parameters as follows. There are two independent coupling constants $g$ and $m=gc$ with the latter identified with the Romans mass in massive type IIA theory. In the following analysis, we will choose the value of $c=1$ and also set $g=m=1$. In this case, the scalar potential can be obtained as follows
\begin{eqnarray} 
 V&=&\frac{1}{2}e^{-\varphi-2\phi_1-2\phi_2-2\phi_3}\left[e^{2(\phi_1+\phi_2+\phi_3)}\left[2e^{2(\phi_2+\phi_3)}(\zeta_1^2+4e^{2\varphi}\zeta_2^2\zeta_3^2) -8(\zeta_1^2+\zeta_2^2+\zeta_3^2)\times\right.\right.
 \nonumber \\
 & &\times e^{\varphi+\phi_1+\phi_2+\phi_3}+2e^{2\phi_3}\zeta_2^2+e^{2\phi_1}\left[e^{2(\phi_2+\phi_3)}\zeta_1^4-8e^{2(\varphi+\phi_2+\phi_3)}\zeta_1\zeta_2\zeta_3+e^{2\phi_2}(2\zeta_3^2 \right.\nonumber \\ 
& &\left. \left.+e^{2\phi_3}(\zeta_2^2+\zeta^2)^2)+2\zeta_1^2\{e^{2\phi_3}(4e^{2\varphi}+e^{2\phi_2})\zeta_3^2+e^{2\phi_2}\zeta_2^2(4e^{2\varphi}+e^{2\phi_3}+8e^{2(\varphi+\phi_3)}\zeta_3^2)\} \right]  \right] \nonumber \\
& &+e^{4(\phi_1+\phi_2)}-8e^{\varphi+3(\phi_1+\phi_2)+\phi_3}-8e^{\varphi+\phi_1+\phi_2+3\phi_3}(e^{2\phi_1}+e^{2\phi_2})
+2\zeta_3^2\chi^2e^{4(\phi_1+\phi_2)+2\phi_3+2\varphi} 
\nonumber \\
& &-2e^{2(\phi_1+\phi_2+\phi_3)}(4e^{2\varphi}+e^{2\phi_1}+e^{2\phi_2})+e^{4\phi_3}[e^{2(\varphi+2\phi_1+2\phi_2)}+(e^{2\phi_1}-e^{2\phi_2})^2]\nonumber \\
& &+2\chi e^{2(\varphi+\phi_1+\phi_2+\phi_3)}\left[e^{2(\phi_1+\phi_2+\phi_3)}(4\zeta_1^3\zeta_2\zeta_3-\zeta_1^2-\zeta_2^2-\zeta_3^2) \right. \nonumber \\
& & \left. +4\zeta_1\zeta_2\zeta_3\left[e^{2(\phi_1+\phi_2)}+e^{2(\phi_1+\phi_3)}+e^{2(\phi_2+\phi_3)}+e^{2(\phi_1+\phi_2+\phi_3)}(\zeta_2^2+\zeta_3^2)\right]\right] \nonumber \\ 
& &+e^{2\varphi}\chi^2\left[e^{4(\phi_1+\phi_2)}+e^{4\phi_3}(e^{2\phi_1}-e^{2\phi_2})^2-2e^{2(\phi_1+\phi_2+\phi_3)}(e^{2\phi_1}+e^{2\phi_2})\right.\nonumber \\
& &\left. \left. +e^{2(\phi_1+\phi_2+2\phi_3)}\left[2e^{2\phi_2}\zeta_1^2+e^{2\phi_1}(2\zeta_2^2+e^{2\phi_2}(\zeta_1^2+\zeta_2^2+\zeta_3^2)^2)\right]\right] \right]. \label{Poten}
\end{eqnarray}  
This scalar potential admits four supersymmetric $AdS_4$ vacua as collected in table \ref{table1} with $V_0$ denoting the corresponding cosmological constants.   

\begin{table}[h]
\centering
\begin{tabular}{|c|c|c|c|}
\hline
Super- & Residual & $(\zeta_1,\zeta_2,\zeta_3,\chi)$ & $V_0$ \\
symmetry & symmetry  & $(\phi_1,\phi_2,\phi_3,\varphi)$ & \\
\hline
$N=1$ & $G_2$ & $\left(-\frac{1}{4 (2^{\frac{1}{3}})},-\frac{1}{4 (2^{\frac{1}{3}})},-\frac{1}{4 (2^{\frac{1}{3}})},-\frac{1}{4 (2^{\frac{1}{3}})}\right)$ & $-\frac{512 (2^{\frac{1}{3}})}{25}\sqrt{\frac{3}{5}}$ \\ 
& & $\left(-\ln \left[\frac{\sqrt{15}}{4 (2^{\frac{1}{3}})}\right],-\ln \left[\frac{\sqrt{15}}{4 (2^{\frac{1}{3}})}\right],-\ln \left[\frac{\sqrt{15}}{4 (2^{\frac{1}{3}})}\right],-\ln \left[\frac{\sqrt{15}}{4 (2^{\frac{1}{3}})}\right]\right)$ &\\ \hline
$N=1$ & $SU(3)$ & $\left(-\frac{\sqrt{3}}{4},-\frac{\sqrt{3}}{4},\frac{1}{4},\frac{1}{4}\right)$ & $-\frac{768}{25}\sqrt{\frac{3}{5}}$ \\ 
& & $\left(-\ln \left[\frac{\sqrt{5}}{4}\right],-\ln \left[\frac{\sqrt{5}}{4}\right],-\ln \left[\frac{\sqrt{15}}{4}\right],-\ln \left[\frac{\sqrt{15}}{4}\right]\right)$ &\\ \hline
$N=2$ & $U(3)$ & $\left(0,0,-\frac{1}{2},-\frac{1}{2}\right)$ & $-12\sqrt{3}$ \\ 
& & $\left(\ln\sqrt{2} ,\ln\sqrt{2},-\ln \left[\frac{\sqrt{3}}{2}\right],-\ln \left[\frac{\sqrt{3}}{2}\right]\right)$ &\\ \hline
$N=3$ & $SO(4)$ & $\left(\frac{1}{2 (2^{\frac{1}{3}})},-\frac{1}{2 (2^{\frac{1}{3}})},-\frac{1}{2 (2^{\frac{1}{3}})},-\frac{1}{ 2^{\frac{1}{3}}}\right)$ & $-\frac{32 (2^{\frac{1}{3}})}{\sqrt{3}}$ \\ 
& & $\left(-\ln \left[\frac{\sqrt{3}}{2 (2^{\frac{1}{3}})}\right],-\ln \left[\frac{\sqrt{3}}{2 (2^{\frac{1}{3}})}\right],-\ln \left[\frac{\sqrt{3}}{2 (2^{\frac{1}{3}})}\right],-\ln \left[\frac{\sqrt{3}}{ 2^{\frac{1}{3}}}\right]\right)$ &\\ \hline
\end{tabular}
\caption{Supersymmetic $AdS_4$ vacua of dyonic $ISO(7)$ gauged supergravity in $SO(3)$ invariant sector with $g=m=1$.}\label{table1}
\end{table}      

\subsection{$N=1$ formulation and BPS equations}
Although the $SO(3)$ invariant sector of the $ISO(7)$ gauged supergravity is described by $N=4$ gauged supergravity coupled to three vector multiplets, the supersymmetric Janus solutions of interest here can be found within the $N=1$ subtruncation of the $ISO(3)\times SO(3)$ gauged $N=4$ supergravity. This is similar to the study of holographic RG flows performed in \cite{ISO7_N3_flow}. The $N=1$ subtruncation is effectively $N=1$ supergravity coupled to four chiral multiplets with four complex scalars given by
\begin{eqnarray}
z_1&=&-\zeta_1+ie^{-\phi_1},\qquad z_2=-\zeta_2+ie^{-\phi_2},\nonumber \\
z_3&=&-\zeta_3+ie^{-\phi_3},\qquad z_4=-\chi+ie^{-\varphi}\, .
\end{eqnarray}
The corresponding bosonic $N=1$ Lagrangian can be written as
\begin{equation}
e^{-1}\mc{L}_{N=1}=\frac{1}{2}R-\frac{1}{2}K_{I\bar{J}}\pd_\mu z^I\pd^\mu \bar{z}^{\bar{J}}-V
\end{equation}
with the Kahler metric 
\begin{equation}
K_{I\bar{J}}dz^Id\bar{z}^{\bar{J}}=-\sum_{i=1}^3\frac{2}{(z_i-\bar{z}_i)^2}dz_id\bar{z}_i-\frac{1}{(z_4-\bar{z}_4)^2}dz_4d\bar{z}_4
\end{equation}
obtained from the Kahler potential of the form
\begin{equation}
K=-2\sum_{i=1}^3\ln[-i(z_i-\bar{z}_i)]-\ln[-i(z_4-\bar{z}_4)].
\end{equation}
The scalar potential is given in \eqref{Poten} and can be written in terms of the superpotential as
\begin{equation}
V=8K^{I\bar{J}}\pd_{z^I}\mc{W}\pd_{\bar{z}^{\bar{J}}}\mc{W}-12\mc{W}^2\, .
\end{equation}
The superpotential is in turn given by
\begin{equation}
\mc{W}=\frac{1}{2}e^{\frac{K}{2}}\sqrt{\mc{Z}\overline{\mc{Z}}}
\end{equation}
with 
\begin{equation}
\mc{Z}=2+8z_1z_2z_3+2z_4(z_1^2+z_2^2+z_3^2).
\end{equation}
\indent In terms of the eight real scalars $(\zeta_1,\zeta_2,\zeta_3,\chi,\phi_1,\phi_2,\phi_3,\varphi)$, the ``real'' superpotential $W=|\mc{W}|$ is given by
\begin{eqnarray}
W=\frac{1}{8}e^{\frac{\varphi}{2}+\phi_1+\phi_2+\phi_3}\sqrt{W_r^2+W_i^2}
\end{eqnarray}
for 
\begin{eqnarray}
W_r&=&8e^{-\phi_3}(\zeta_1\zeta_2-e^{-\phi_1-\phi_2})+2e^{-\varphi}(\zeta_1^2+\zeta_2^2+\zeta_3^2-e^{-2\phi_1}-e^{-2\phi_2}-e^{-2\phi_3})\nonumber \\
& &+8\zeta_3(e^{-\phi_2}\zeta_1+\zeta_2e^{-\phi_1})+4\chi(e^{-\phi_1}\zeta_1+e^{-2\phi_2}\zeta_2+e^{-\phi_3}\zeta_3),\\
W_i&=&2+8e^{-\phi_3}(e^{-\phi_2}\zeta_1+e^{-\phi_1}\zeta_2)+2\chi(e^{-2\phi_1}+e^{-2\phi_2}+e^{-2\phi_3}-\zeta_1^2-\zeta_2^2-\zeta_3^2)\nonumber \\
& &+8\zeta_3(e^{-\phi_1-\phi_2}-\zeta_1\zeta_2)+4e^{-\varphi}(e^{-\phi_1}\zeta_1+e^{-\phi_2}\zeta_2+e^{-\phi_3}\zeta_3).
\end{eqnarray}
The scalar potential can be written in terms of $W$ as
\begin{equation}
V=4\sum_{i=1}^3\left[e^{-2\phi_i}\left(\frac{\pd W}{\pd \zeta_i}\right)^2+\left(\frac{\pd W}{\pd \phi_i}\right)^2\right]+8e^{-2\varphi}\left(\frac{\pd W}{\pd \chi}\right)^2+8\left(\frac{\pd W}{\pd \varphi}\right)^2-6W^2\, .
\end{equation}
\indent We now consider the fermionic supersymmetry transformations and derive the corresponding BPS equations for supersymmetric Janus solutions. For four-dimensional gauged supergravity, the procedure has been initiated in \cite{warner_Janus} to which the reader is referred for more detail. We will mostly present the final result with a brief review of relevant formulae. The metric ansatz takes the form of an $AdS_3$-sliced domain wall
\begin{equation}
ds^2=e^{2A(r)}\left(e^{\frac{2\rho}{\ell}}dx^2_{1,1}+d\rho^2\right)+dr^2\,
.
\end{equation}
This gives the metric of standard domain walls in the limit $\ell\rightarrow \infty$. With the vielbein components
\begin{equation}
e^{\hat{\alpha}}=e^{A+\frac{\rho}{\ell}}dx^\alpha,\qquad
e^{\hat{\rho}}=e^{A}d\rho,\qquad e^{\hat{r}}=dr,
\end{equation}
it is straightforward to compute the following non-vanishing components of the spin connection 
\begin{equation}
\omega^{\hat{\rho}}_{\phantom{\hat{\rho}}\hat{r}}=A'e^{\hat{\rho}},\qquad
\omega^{\hat{\alpha}}_{\phantom{\hat{\rho}}\hat{\rho}}=\frac{1}{\ell}e^{-A}e^{\hat{\alpha}},\qquad
\omega^{\hat{\alpha}}_{\phantom{\hat{\rho}}\hat{r}}=A'e^{\hat{\alpha}}
\end{equation}
where $'$ denotes the $r$-derivative. Indices $\alpha,\beta$ and $\hat{\alpha},\hat{\beta}$ take values $0,1$. All the scalars are functions of only $r$. In addition, we will set $\ell=1$ in subsequent numerical analysis given in the next sections.   
\\ 
\indent By the same analysis as in \cite{warner_Janus}, we find that the supersymmetry transformations $\delta\psi^{I}_{\hat{\alpha}}$ give the following equation
\begin{equation}
A'^2=W^2-\frac{1}{\ell^2}e^{-2A}\, .\label{dPsi_BPS_eq}
\end{equation}
The variation $\delta \psi^I_{\hat{\rho}}$ gives the Killing spinor of the form
\begin{equation}
\epsilon=e^{\frac{\rho}{2\ell}}\tilde{\epsilon} 
\end{equation}
for a $\rho$-independent spinor $\tilde{\epsilon}$. 
\\
\indent We also need to impose the projectors of the form
\begin{equation}
\gamma_{\hat{r}}\epsilon=e^{i\Lambda}\epsilon^*\label{gamma_r_pro}
\end{equation}
and
\begin{equation}
\gamma_{\hat{\rho}}\epsilon=i\kappa e^{i\Lambda}\epsilon^*\label{gamma_rho_pro}
\end{equation}
with $\kappa^2=1$ and an $r$-dependent phase $\Lambda$. We note that there can be more than one Killing spinor if the solutions preserve more supersymmetry. In this case, Killing spinors in different representations of the residual symmetry can satisfy the projectors \eqref{gamma_r_pro} and \eqref{gamma_rho_pro} with different phases.
\\
\indent It should also be remarked that we use Majorana representation with $\gamma^{\hat{\mu}}$ real and $\gamma_5$ purely imaginary. The two chiralities of a given spinor are then related to each other by a complex conjugate. The constant $\kappa$ is related to the chirality of the Killing spinor on the two-dimensional defects dual to the $AdS_3$ slices. The explicit form of the Killing spinor is determined to be
\begin{equation}
\epsilon=e^{\frac{A}{2}+\frac{\rho}{2\ell}+i\frac{\Lambda}{2}}\varepsilon^{(0)}
\end{equation}
with possibly an $r$-dependent phase for the spinor $\varepsilon^{(0)}$ satisfying
\begin{equation}
\gamma_{\hat{r}}\varepsilon^{(0)}=\varepsilon^{(0)*}\qquad
\textrm{and}\qquad
\gamma_{\hat{\rho}}\varepsilon^{(0)}=i\kappa\varepsilon^{(0)*}\,
.
\end{equation}
With all these results, we can also find the explicit form of the phase from $\delta \psi^I_{\hat{\alpha}}$ equations
\begin{equation}
e^{i\Lambda}=\frac{\mc{W}}{A'+\frac{i\kappa}{\ell}e^{-A}}\,
.\label{complex_W_phase}
\end{equation}
Using the projector \eqref{gamma_r_pro}, we can determine all the BPS equations for scalars from the variation $\delta\chi^{IJK}$. 
\\
\indent In terms of the $N=1$ truncation, the resulting BPS equations can be written as
\begin{eqnarray}
\zeta_1'&=&-2\frac{A'}{W}e^{-2\phi_1}\frac{\pd W}{\pd \zeta_1}+2e^{-\phi_1}\frac{\kappa e^{-A}}{W\ell}\frac{\pd W}{\pd \phi_1},\label{eq1}\\
\phi_1'&=&-2\frac{A'}{W}\frac{\pd W}{\pd \phi_1}-2\frac{\kappa e^{-A}}{W\ell}e^{-\phi_1}\frac{\pd W}{\pd \zeta_1},\\
\zeta_2'&=&-2\frac{A'}{W}e^{-2\phi_2}\frac{\pd W}{\pd \zeta_2}+2e^{-\phi_2}\frac{\kappa e^{-A}}{W\ell}\frac{\pd W}{\pd \phi_2},\\
\phi_2'&=&-2\frac{A'}{W}\frac{\pd W}{\pd \phi_2}-2\frac{\kappa e^{-A}}{W\ell}e^{-\phi_2}\frac{\pd W}{\pd \zeta_2},\\
\zeta_3'&=&-2\frac{A'}{W}e^{-2\phi_3}\frac{\pd W}{\pd \zeta_3}+2e^{-\phi_3}\frac{\kappa e^{-A}}{W\ell}\frac{\pd W}{\pd \phi_3},\\
\phi_3'&=&-2\frac{A'}{W}\frac{\pd W}{\pd \phi_3}-2\frac{\kappa e^{-A}}{W\ell}e^{-\phi_3}\frac{\pd W}{\pd \zeta_3},\\
\chi'&=&-4\frac{A'}{W}e^{-2\varphi}\frac{\pd W}{\pd \chi}+4e^{-\varphi}\frac{\kappa e^{-A}}{W\ell}\frac{\pd W}{\pd \varphi},\\
\varphi'&=&-4\frac{A'}{W}\frac{\pd W}{\pd \varphi}-4\frac{\kappa e^{-A}}{W\ell}e^{-\varphi}\frac{\pd W}{\pd \chi}\, .\label{eq2}
\end{eqnarray}
The explicit form of these equations is highly complicated, so we refrain from giving them here. It should be noted that these equations reduce to those of the RG flows studied in \cite{ISO7_N3_flow} in the limit $\ell\rightarrow \infty$. 
\\
\indent Finding the BPS equations for supersymmetric Janus solutions involves solving equation \eqref{dPsi_BPS_eq} with a branch cut. We will follow the procedure carried out in \cite{warner_Janus} to work with a smooth numerical analysis. This is achieved by solving the second-order field equations by fixing a turning point of $A$ such that $A'(r_0)=0$ for particular values of scalar fields. Without loss of generality, we will conveniently choose $r_0=0$. We then determine the values of $A(0)$, $\zeta_i'(0)$, $\phi_i'(0)$, $\chi'(0)$ and $\varphi'(0)$ using the BPS equations giving rise to a full set of initial conditions to solve the second order field equations. We  finally check whether the resulting solution satisfies the BPS equations. For completeness, we note all the field equations here 
\begin{eqnarray}
0&=&2A''+3{A'}^2+\frac{e^{-2A}}{\ell^2}+\frac{1}{4}V+\frac{1}{2}\sum_{i=1}^3\left(e^{2\phi_i}{\zeta_i'}^2+{\phi_i'}^2\right)+\frac{1}{2}({\varphi'}^2+e^{2\varphi}{\chi'}^2),\quad\\
0&=&2e^{-3A}\frac{d}{dr}\left(e^{3A}\phi'_i\right)-2{\zeta'_i}^2e^{2\phi_i}-\frac{\pd V}{\pd \phi_i},\qquad i=1,2,3,\\
0&=&e^{-3A}\frac{d}{dr}\left(e^{3A}\varphi'\right)-e^{2\varphi}{\chi'}^2-\frac{\pd V}{\pd \varphi},\\
0&=&2e^{-3A}\frac{d}{dr}\left(e^{3A}e^{2\phi_i}\zeta_i'\right)-\frac{\pd V}{\pd \zeta_i},\qquad i=1,2,3,\\
0&=&e^{-3A}\frac{d}{dr}\left(e^{3A}e^{2\varphi}\chi'\right)-\frac{\pd V}{\pd \chi}\, .
\end{eqnarray}
It is straightforward to verify that the BPS equations are compatible with these equations.  

\section{$N=3$ supersymmetric Janus solutions}\label{N3Janus}
\begin{figure}
    \includegraphics[width=1.0\linewidth]{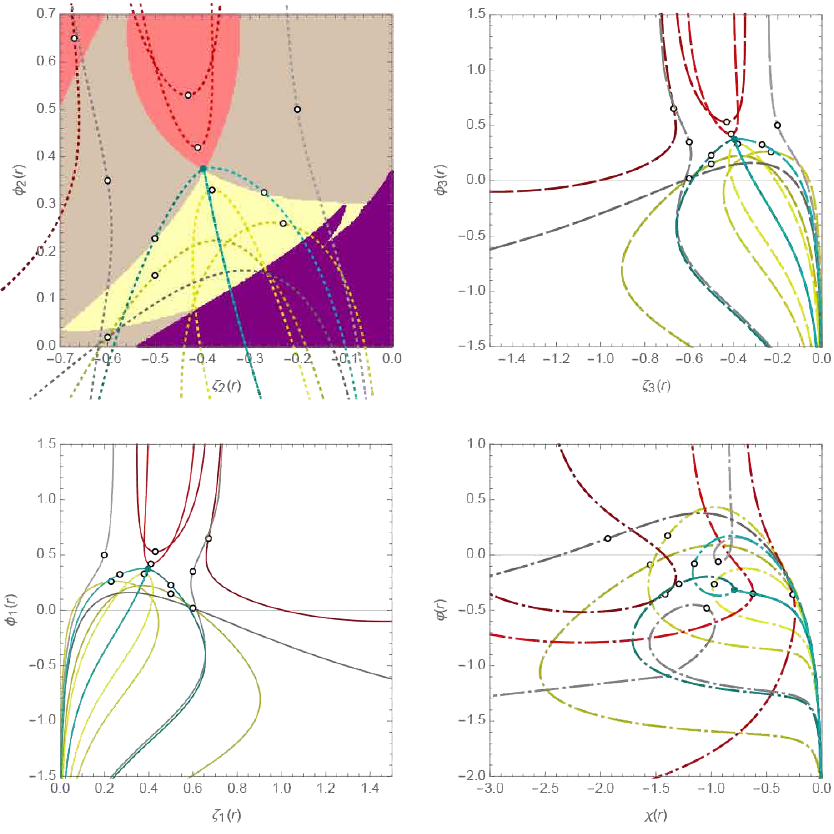}
  \caption{The space of solutions for $N=3$ supersymmetric Janus solutions with $SO(4)$ symmetry. The green dot represents the $N=3$ supersymmetric $AdS_4$ vacuum while the open dots denote turning points at which $A'=0$.} 
  \label{space_N3}
\end{figure}

In this section, we consider supersymmetric Janus solutions preserving $N=3$ supersymmetry. This can be achieved by setting
\begin{equation}
z_3=z_2=-\bar{z}_1
\end{equation}
or equivalently
\begin{equation}
\phi_3=\phi_2=\phi_1\qquad \textrm{and}\qquad \zeta_3=\zeta_2=-\zeta_1\, .
\end{equation}
These relations give rise to the superpotential
\begin{equation}
W=\frac{1}{8}e^{2\phi_1+\frac{1}{2}\varphi}\sqrt{w_r^2+w_i^2}
\end{equation} 
with
\begin{eqnarray}
w_r&=&2+4e^{-2\phi_1}(2+e^{\phi_1-\varphi})\zeta_1+8\zeta^3_1+6\chi (e^{-2\phi_1}-\zeta^2_1),\\
w_i&=&4\chi\zeta_1 e^{-\phi_1}+\zeta^2_1(6e^{-\varphi}-8e^{-\phi_1})-2e^{-3\phi_1}(4+3e^{\phi_1-\varphi}).
\end{eqnarray}
It can also be straightforwardly verified that this truncation leads to a three-fold degenerate eigenvalue of $A_1^{IJ}$ giving rise to this superpotential. The scalar potential is given by 
\begin{eqnarray}
V&=&\frac{1}{2}e^{-\varphi}\left[\left[6e^{4\phi_1}\zeta_1^2(\chi-2\zeta_1)^2+e^{6\phi_1}(1+4\zeta_1^3-3\zeta_1^2\chi)^2 -3\chi^2e^{2\phi_1}\right] \right. \nonumber \\
& &\left. -8e^{2\varphi}-3e^{2\phi_1}+6e^{4\phi_1}\zeta_1^2+9e^{6\phi_1}\zeta_1^4-24e^{\phi_1+\varphi}(1+e^{2\phi_1}\zeta_1^2)\right]
\end{eqnarray}
which agrees precisely with that given in \cite{ISO7_Guarino} up to field redefinitions.  
\\
\indent At this stage, we are left with four scalars $\varphi$, $\chi$, one of $\phi_i$ and one of $\zeta_i$, $i=1,2,3$. However, using the above truncation in the BPS equations \eqref{eq1} to \eqref{eq2} results in the following two consistency conditions
\begin{eqnarray}
& &A'e^{A+\phi_1}\left[e^{2\phi_1+\varphi}[e^{2\phi_1}(2\zeta_1^3-1)-6\zeta_1]-2\zeta_1 e^{2\phi_1}\right.\nonumber \\
& &\left.+2e^{2\varphi}[5\chi+e^{2\phi_1}(1+\zeta_1\chi(\zeta_1-\chi))]\right]\nonumber \\
& &\qquad =-e^{2(\varphi+\phi_1)}(2\zeta_1-\chi)[8\zeta_1+3\chi+e^{2\phi_1}(1+4\zeta_1^3-3\zeta_1^2\chi)]-8e^{2\varphi}\nonumber \\
&& \qquad \phantom{=}+3e^{2\phi_1}(1-e^{2\phi_1}\zeta_1^2+2e^{\phi_1+\varphi}(3e^{2\phi_1}\zeta_1^2-1)\label{con1}
\end{eqnarray}
and
\begin{eqnarray}
& & A'e^A\left[4(e^{2\phi_1}\zeta_1+e^{2\varphi}\chi)+4e^{2(\varphi+\phi_1)}[1+\zeta_1\chi(\chi-3\zeta_1)]\right.\nonumber \\
& &\left.+2e^{\phi_1+\varphi}[2\zeta_1+e^{2\phi_1}(1+10\zeta_1^3)]\right]\nonumber \\
& & \qquad =3e^{\phi_1}+4e^{\varphi}+4\zeta_1^2e^{2\phi_1}(3e^\varphi-2e^{\phi_1})+\zeta_1 e^{4\phi_1}[4e^{\varphi}+\zeta_1^3(9e^{\phi_1}-8e^\varphi)]\nonumber \\
& &\qquad \phantom{=}+e^{\phi_1+2\varphi}\left[16\zeta_1^2-4\zeta_1\chi+3\chi^2+e^{4\phi_1}(1+4\zeta_1^3-3\zeta_1^2\chi)^2\right] \nonumber \\
& &\qquad \phantom{=}+4e^{3\phi_1+2\varphi}[\zeta_1+8\zeta_1^4+\chi-\zeta_1^2\chi(3\zeta_1-2\chi)].\label{con2}
\end{eqnarray}
We note here that for holographic RG flows with $\ell\rightarrow \infty$, these conditions are identically satisfied. Therefore, there are four independent scalars in the case of RG flows while Janus solutions are characterized only by two independent scalars. It could also be interesting to study holographic RG flow solutions in this truncation which have not been considered in previous works. In principle, we can solve the conditions \eqref{con1} and \eqref{con2} to express two of the four scalars in terms of the remaining two and solve the BPS equations for these two independent scalars and $A(r)$. We now perform a numerical analysis to find possible supersymmetric Janus solutions. 
\\
\indent Following \cite{warner_Janus}, we first determine the space of solutions by considering various positions of the $A(r)$ turning point in the field space. The result from a numerical scan is given in figure \ref{space_N3}. The open circles refer to turning points of $A$ at which $A'=0$ while the green dot represents $N=3$ $AdS_4$ critical point with $SO(4)$ symmetry. We have used the same color codes as in \cite{warner_Janus} for various different regions. It should be remarked again that among the eight scalars, only two of them are independent. Therefore, we have chosen to represent the shaded regions only in $(\phi_2,\zeta_2)$ on which the structure of the space of solutions is less involved. The other three planes are given for representing the full Janus solutions for all scalars. Although only two scalars are independent, we give the solutions with all scalars for the purpose of comparison with other solutions to be given in subsequent sections.  
\\
\indent The generic structure of possible solutions is very similar to supersymmetric Janus solutions from the maximal $SO(8)$ gauged supergravity studied in \cite{warner_Janus} with a major difference being the maximally supersymmetric $AdS_4$ vacuum with $SO(8)$ symmetry on M2-branes replaced by the $N=8$ SYM phase on D2-branes. The latter corresponds to $\varphi\sim \phi_i\rightarrow -\infty$ and $\chi\sim \zeta_i\rightarrow 0$. If the turning points lie on the yellow region, we find Janus solutions between $N=8$ SYM phases similar to the solutions found in \cite{Minwoo_4DN8_Janus} for $N=1$ Janus solutions. These solutions should describe two-dimensional conformal defects or interfaces within $N=8$ SYM in three dimensions. 
\\
\indent If the turning points lie closer to the boundaries between the yellow and grey regions, the solutions still interpolate between $N=8$ SYM phases but approach the $SO(4)$ $AdS_4$ critical point on one side before flowing to the SYM phase. These are again similar to the Janus solutions in \cite{warner_Janus}. In addition, if the turning points lie very close to the boundary, the solutions will approach the $SO(4)$ $AdS_4$ critical point arbitrarily close and stay at the critical point for an arbitrarily long period of time. By a similar argument as in \cite{warner_Janus}, we expect these solutions to describe conformal interfaces between the $SO(4)$ critical point on one side and the SYM phase on the other side, denoted by SYM/$SO(4)$ Janus solution. The $N=3$ $SO(4)$ conformal phase on one side is generated by the usual RG flows from the $N=8$ SYM phase. 

\begin{figure}
  \centering
  \begin{subfigure}[b]{1.0\linewidth}
    \includegraphics[width=\linewidth]{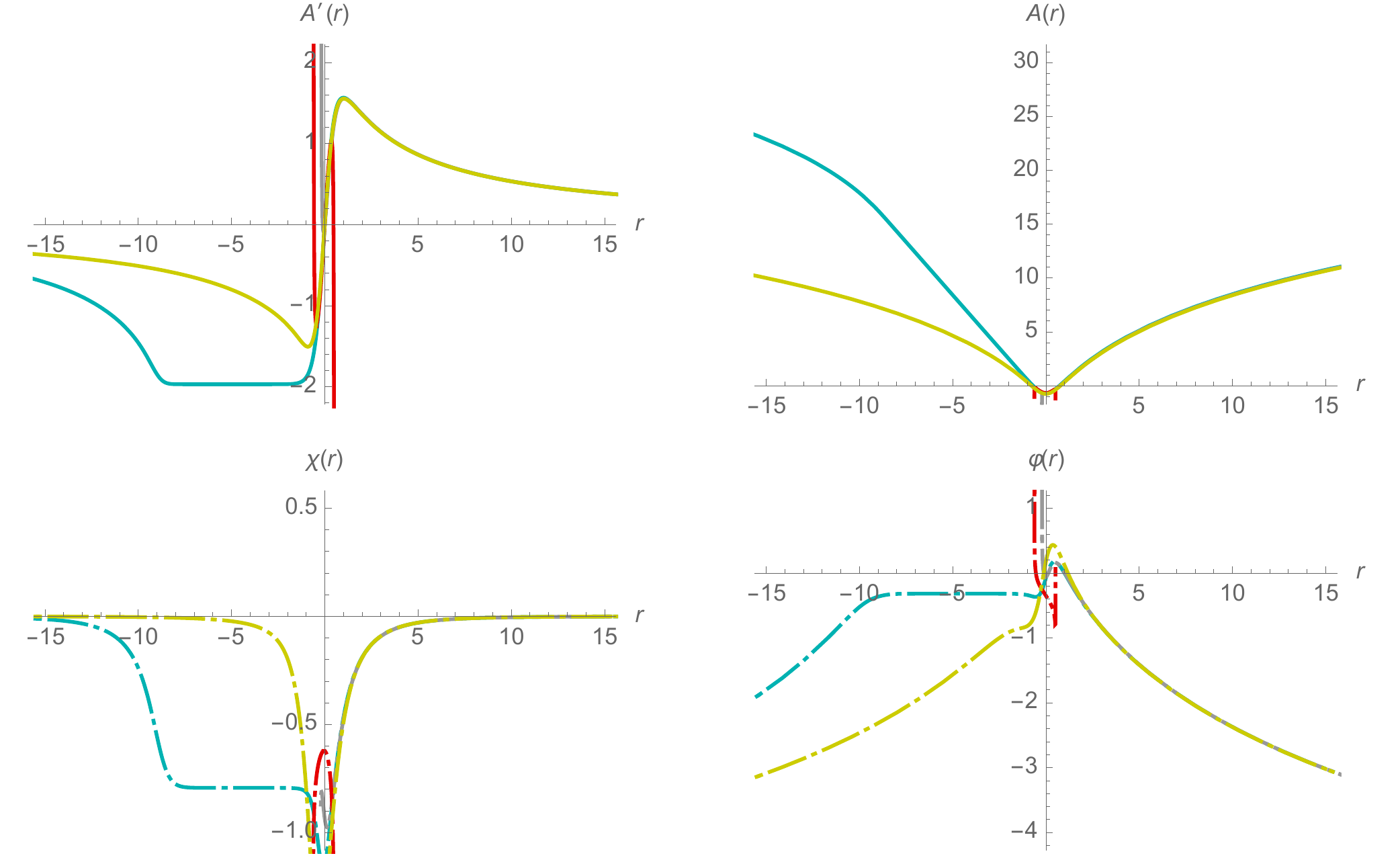}
  \end{subfigure}\\
  \begin{subfigure}[b]{1.0\linewidth}
    \includegraphics[width=\linewidth]{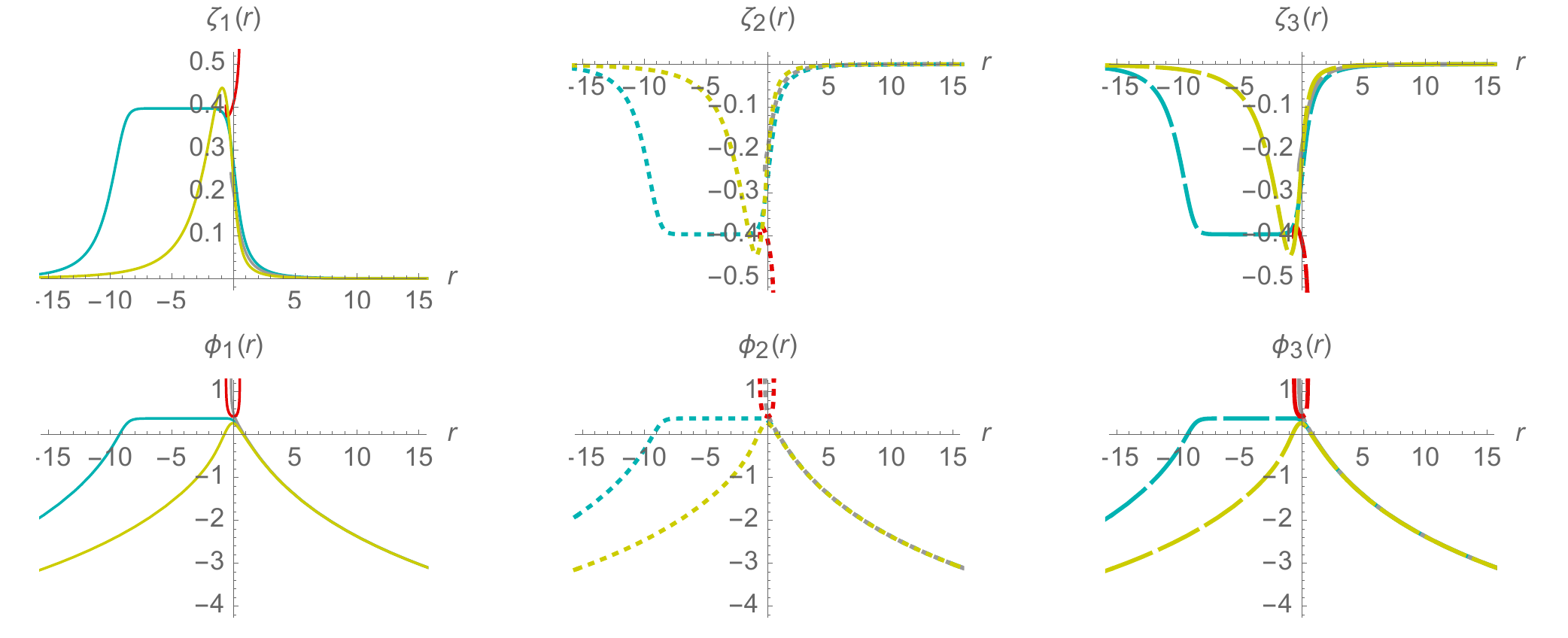}
  \end{subfigure}
  \caption{Examples of supersymmetric Janus solutions interpolating between $N=8$ SYM (yellow), $N=8$ SYM and $SO(4)$ $AdS_4$ (light blue), $N=8$ SYM and a singulary (grey), and between singularities (red).} 
  \label{N3_profile}
\end{figure}

Solutions with turning points on the grey regions interpolate between $N=8$ SYM and singular geometries, SYM/singularity Janus solutions. On the other hand, there are solutions interpolating between singular geometries on both sides if the turning points lie on the red region, singularity/singularity or simply singular Janus solutions. These solutions would decribe boundary SCFTs as pointed out in \cite{ICFT_BCFT}. However, it turns out that all singularities arising from these solutions are unphysical in both four- and ten-dimensional frameworks. In particular, we have verified that these singularities lead to the scalar potential that is unbounded from above and diverging ten-dimensional metric component $\hat{g}_{00}$. The latter is obtained by the formula given in appendix D of \cite{ISO7_N3_flow}. 
\\
\indent Examples of solutions in all these categories are shown in figure \ref{N3_profile}. From the figure, it is clearly seen that the solution represented by the light blue curve exhibits the $SO(4)$ critical point on the left with constant scalars and $A\sim r$. The conformal interfaces corresponding to all of these solutions preserve $N=(3,0)$ or $N=(0,3)$ supersymmetries on the interfaces depending on the values of $\kappa=1$ or $\kappa=-1$. We also note that in the plots of the singular solution represented by the red lines appears to be cut at certain values of $r$. This arises from the limitation of the numerical computation being used here. Finally, we remark that there are no supersymmetric Janus solutions with the turning points on the purple region. In this case, the constraints \eqref{con1} and \eqref{con2} are not satisfied.

\section{$N=2$ supersymmetric Janus solutions}\label{N2Janus}        
\begin{figure}
    \includegraphics[width=1.0\linewidth]{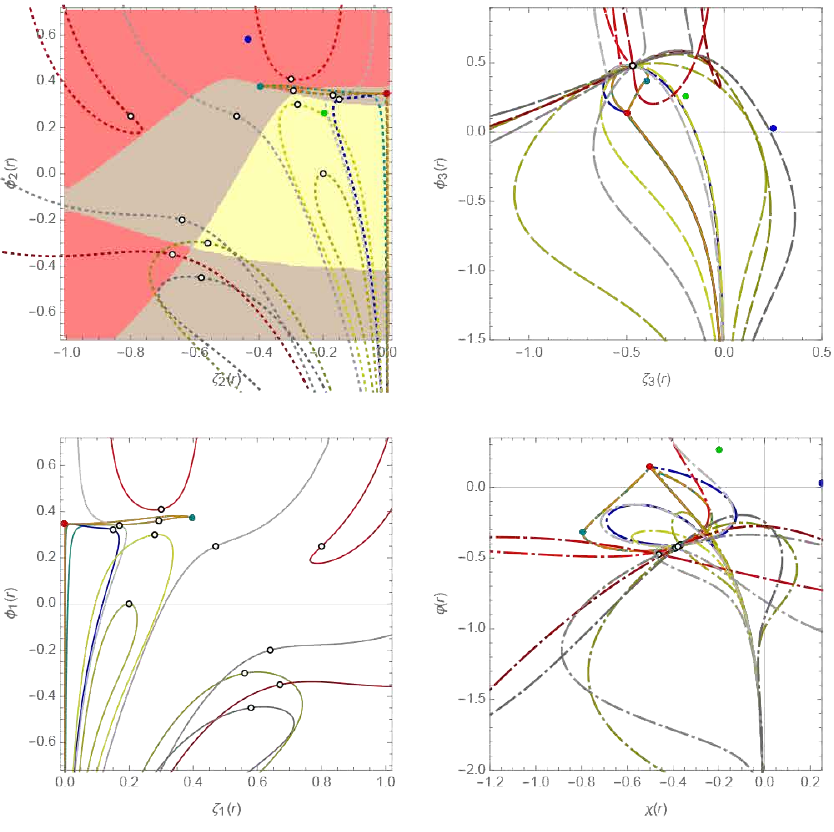}
  \caption{The space of solutions for $N=2$ supersymmetric Janus solutions with $SO(3)\times U(1)$ symmetry. The dark green, red, blue and green dots repectively represent $SO(4)$, $U(3)$, SU(3) and $G_2$ symmetric $AdS_4$ critical points while the open dots denote turning points at which $A'=0$.} 
  \label{N2_space}
\end{figure}

\begin{figure}
  \centering
  \begin{subfigure}[b]{1.0\linewidth}
    \includegraphics[width=\linewidth]{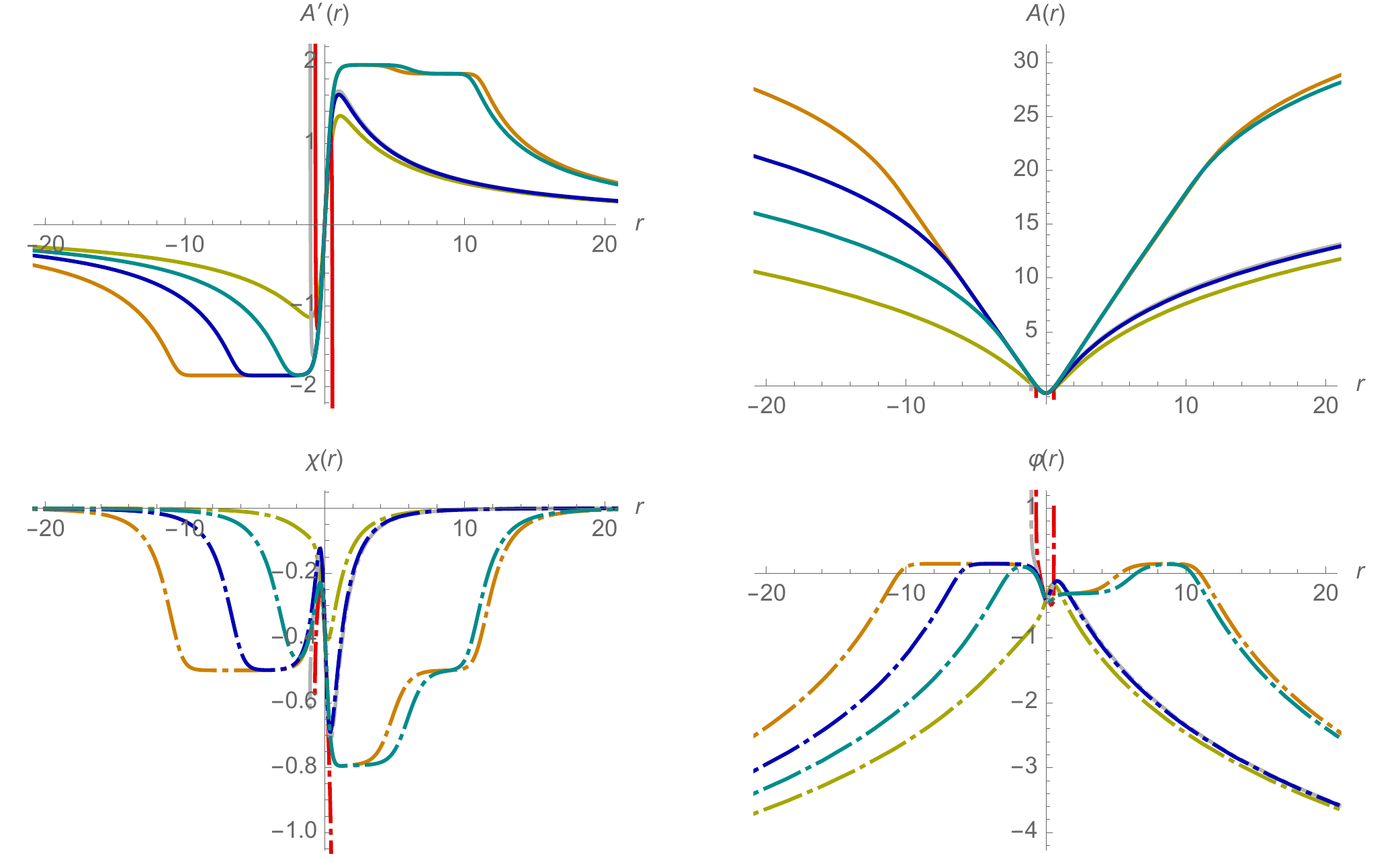}
  \end{subfigure}\\
  \begin{subfigure}[b]{1.0\linewidth}
    \includegraphics[width=\linewidth]{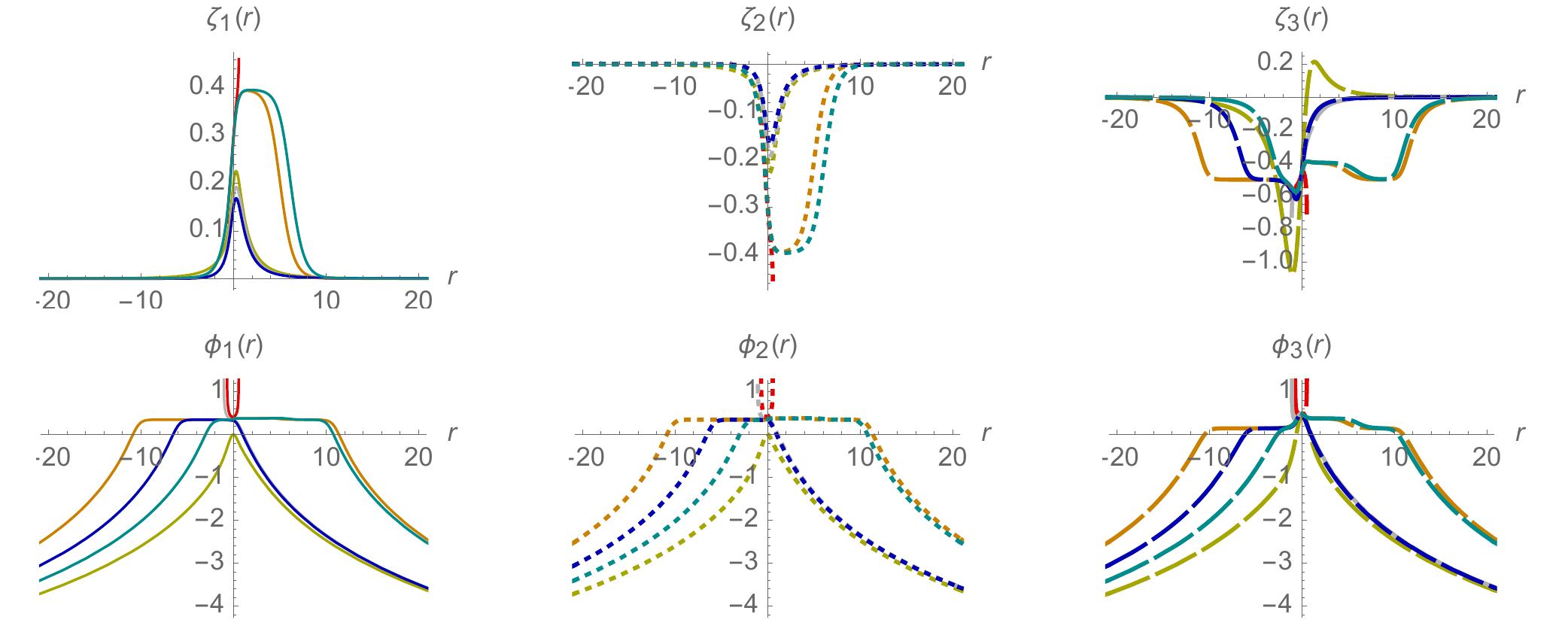}
  \end{subfigure}
  \caption{Examples of supersymmetric Janus solutions interpolating between $N=8$ SYM (yellow) and between SYM and a singularity (grey). The red line represents a singular solution interpolating between two singular geometries on both sides. The solution shown by the orange line describes an $SO(4)/U(3)$ conformal interface within the $N=8$ SYM theory. The blue and green lines correspond to SYM/$U(3)$ and SYM/$SO(4)$ Janus solutions, respectively.} 
  \label{N2_profile}
\end{figure}

In this section, we repeat the same analysis as in $N=3$ supersymmetric Janus solutions in the case of solutions preserving $N=2$ supersymmetry. To achieve this, we first make a truncation as given in \cite{ISO7_N3_flow} by imposing the following condition       
\begin{equation}
z_1=-\bar{z}_2
\end{equation}
which implies
\begin{equation}
\phi_2=\phi_1\qquad \textrm{and}\qquad \zeta_2=-\zeta_1\, .\label{N2_truncation}
        \end{equation}        
In this case, the solutions preserve $SO(3)\times U(1)$ symmetry in the full $N=8$ gauged supergravity. A number of holographic RG flows within this sector have already been studied in \cite{ISO7_N3_flow}. In this paper, we are interested in finding Janus solutions. In this case, the solutions preserve $N=(2,0)$ or $N=(0,2)$ supersymmetries on the interfaces. 
\\
\indent Similar to the $N=3$ solutions given in the previous section, imposing the truncation \eqref{N2_truncation} on the BPS equations results in two consistency conditions. This is similar to the case of holographic RG flows studied in \cite{ISO7_N3_flow} in which the truncation also imposes two additional conditions. These conditions are much more complicated than those of the $N=3$ case, so we refrain from giving them here. With these conditions imposed, we are left with four scalars to describe the Janus solutions. As in the previous sections, we will give numerical solutions for all scalars for comparison with other cases although there are only four independent scalars. By solving the truncated BPS equations subject to the aforementioned two constraints, we find the space of $N=2$ Janus solutions shown in figure \ref{N2_space}. In this case, the space of solutions is four-dimensional. As a manner of representation, we have chosen to consider only a number of projections on two-dimensional planes to make things more manageable. There are four supersymmetric $AdS_4$ vacua appearing on the diagram. These are $SO(4)$, $U(3)$, SU(3) and $G_2$ symmetric critical points with $N=3,2,1,1$ supersymmetries and represented by dark green, red, blue and green dots, repectively. Some examples of various possible solutions are shown in figure \ref{N2_profile}.         
\\
\indent As in the $N=3$ case, there are solutions with turning points on the yellow region that interpolate between $N=8$ SYM phases as shown by the yellow line in figure \ref{N2_profile}. Solutions with turning points on the grey and red regions describe SYM/singularity or singularity/SYM and singularity/singularity Janus solutions, respectively. Examples of these solutions are shown by the grey and red lines in figure \ref{N2_profile}. Most of these solutions have analogues in the $N=3$ case considered in the previous section. However, there is a special solution with the turning point on the boundary between yellow, grey and red regions that interpolates among the SYM and $SO(4)$ or $U(3)$ conformal phases. The solutions describe SYM/$U(3)$ and SYM/$SO(4)$ interfaces are given by the blue and green lines in the figures. The SYM phase on one side of these solutions undergoes an RG flow respectively to $U(3)$ and $SO(4)$ conformal phases.
\\
\indent There is also solution interpolating between SYM phases and both the $U(3)$ and $SO(4)$ critical points. This solution is shown by the orange line in figures \ref{N2_space} and \ref{N2_profile}. In this solution, the turning point lies between the $U(3)$ and $SO(4)$ phases. On the right side, the solution begins at the SYM phase and undergoes a series of RG flows to the $U(3)$ and the $SO(4)$ conformal phases. The solutions then proceeds to the $U(3)$ phase on the left side and is finally back to the SYM phase via another RG flow. We expect this solution to describe an interface between $U(3)$ and $SO(4)$ conformal phases within the $N=8$ SYM. It should be remarked here that no solutions approach $SU(3)$ and $G_2$ conformal phases with $N=1$ supersymmetry. We also point out that on the $(\phi_2,\zeta_2)$ plane, there appear to be solutions passing through the $G_2$ critical point. However, this is not the case as can be seen on the other planes in the figure. Therefore, this is just an artefact of projecting the four-dimensional space of solutions on a particular two-dimensional plane.     
                
\section{$N=1$ supersymmetric Janus solutions}\label{N1Janus}
We now look at Janus solutions with $N=1$ supersymmetry. In this case, there are a number of possibilities which will be separately discussed in the following analysis. Some solutions with $G_2$ and $SU(3)$ symmetries have also been given in \cite{Minwoo_4DN8_Janus}.

\subsection{$N=1$ supersymmetric Janus solutions with $G_2$ symmetry}
\begin{figure}
  \centering
  \begin{subfigure}[b]{0.45\linewidth}
    \includegraphics[width=\linewidth]{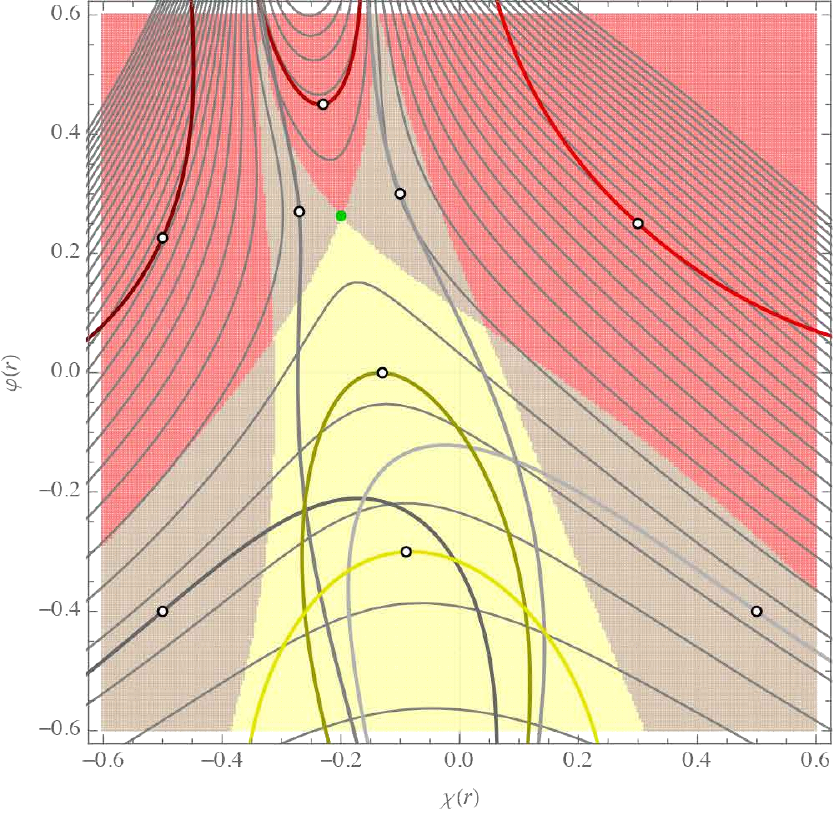}
 \caption{Space of solutions for $G_2$ symmetric Janus with turning points on various different regions.}
  \end{subfigure}\quad
  \begin{subfigure}[b]{0.45\linewidth}
    \includegraphics[width=\linewidth]{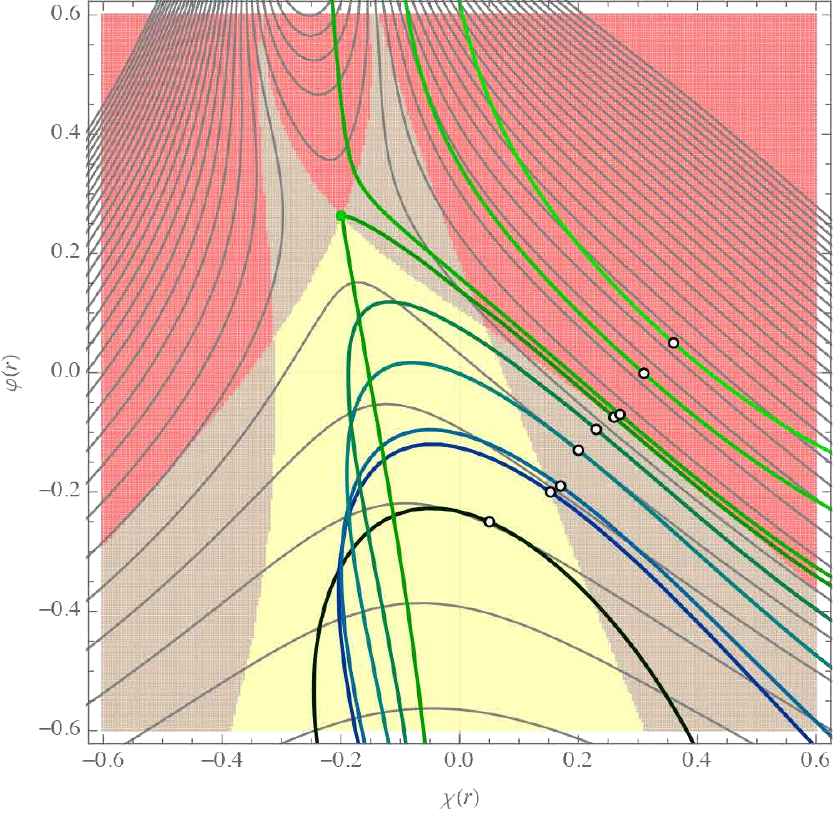}
\caption{Space of solutions for $G_2$ symmetric Janus with turning points moving across boundaries of different regions.}
  \end{subfigure}
  \caption{Space of solutions for generic $G_2$ symmetric Janus solutions. The green dot represents the $G_2$ $AdS_4$ critical point while open dots denote the turning points.} 
  \label{N1_G2_space1}
\end{figure}

\begin{figure}
    \includegraphics[width=1.0\linewidth]{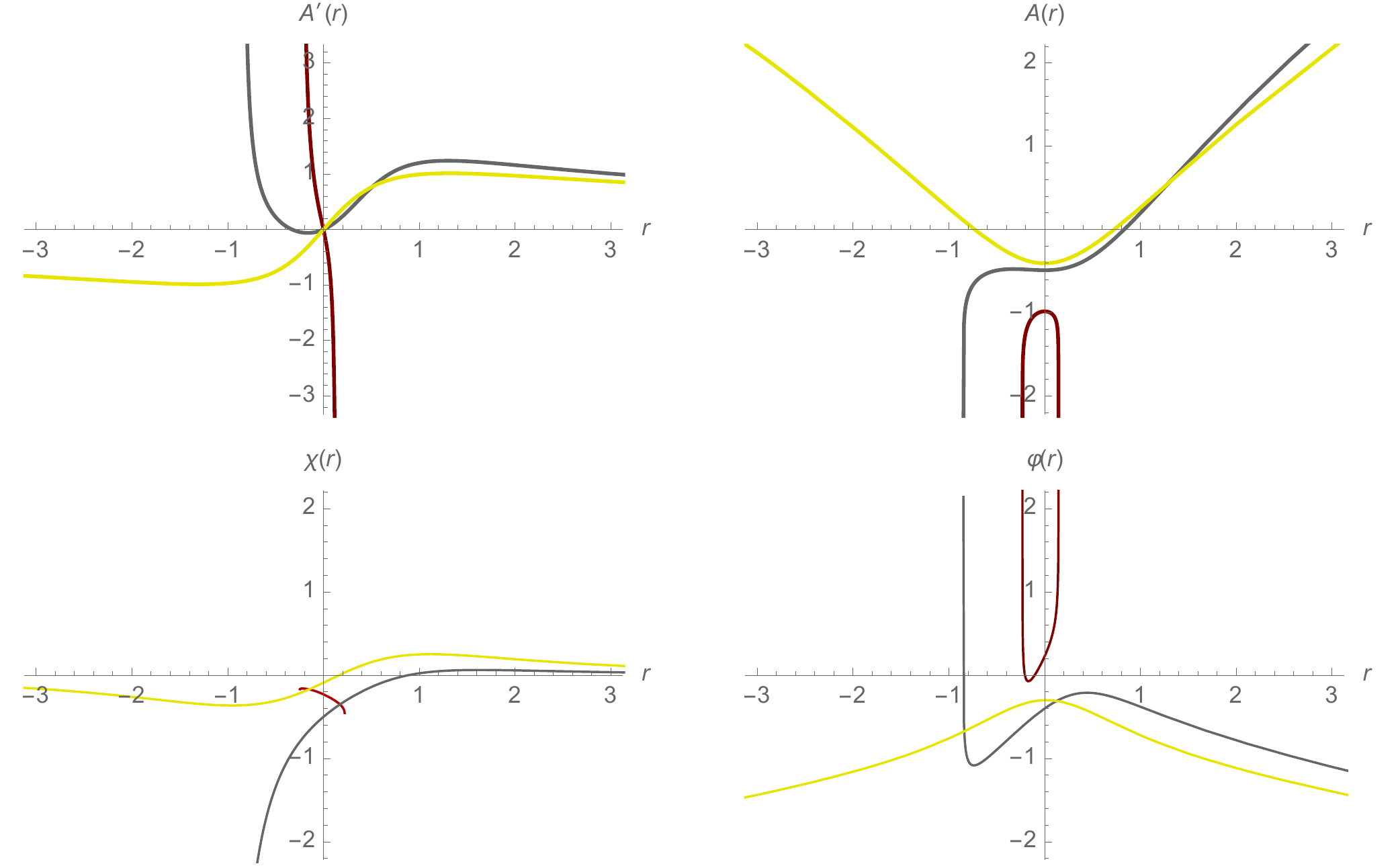}
  \caption{Examples of $N=1$ Janus solutions with $G_2$ symmetry interpolate between SYM phases (yellow), SYM and singularity (grey), and between singularities (dark red).} 
  \label{G2_sol_gen}
\end{figure}

\begin{figure}
    \includegraphics[width=1.0\linewidth]{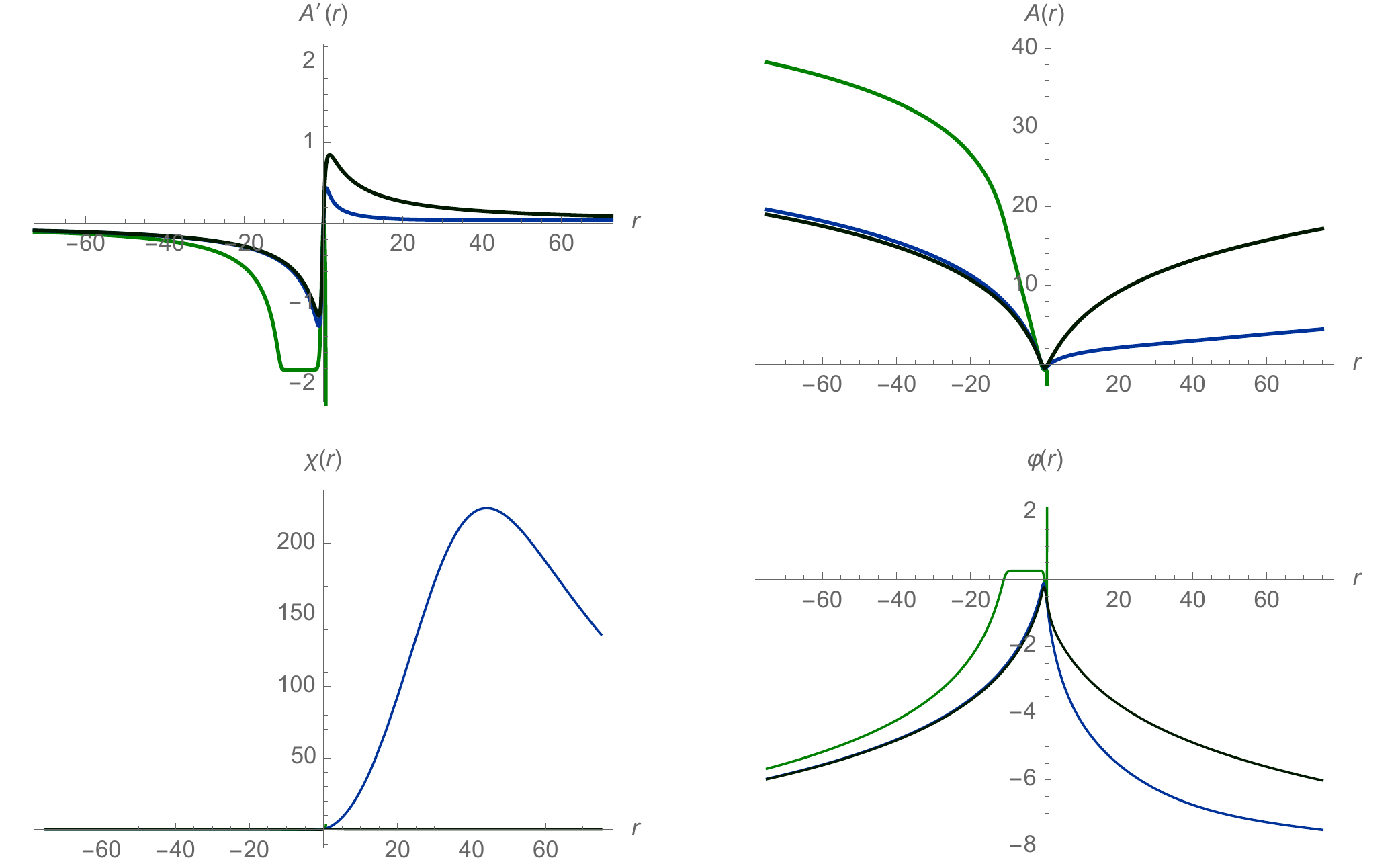}
  \caption{$N=1$ Janus solution with $G_2$ symmetry interpolates between SYM and a singularity with an intermediate $G_2$ conformal phase (green). The solutions shown by black and blue lines describe SYM/SYM interfaces.} 
  \label{G2_sol_gen1}
\end{figure}

We begin with solutions preserving $G_2$ symmetry described by the following relations
\begin{equation}
\zeta_{1}=\zeta_2=\zeta_3=\chi \qquad \text{and} \qquad \phi_1=\phi_2=\phi_3=\varphi\, .
\end{equation}
Accordingly, there are only two independent scalars which we will choose to be $\varphi$ and $\chi$. Some of the solutions in this sector have already been studied in \cite{Minwoo_4DN8_Janus}, so we mainly focus on new aspects of more general $G_2$ symmetric Janus solutions.  
\\
\indent The space of solutions for generic $G_2$ symmetric Janus solutions is shown in figure \ref{N1_G2_space1} on the left panel. The structure is similar to all the previous cases namely there are SYM/SYM, SYM/singularity or singularity/SYM and singularity/singularity Janus solutions for the turning points on yellow, grey and red regions, respectively. Examples of these solutions are given in figure \ref{G2_sol_gen}. As in the $G_2$ symmetric Janus solutions of $SO(8)$ gauged supergravity studied in \cite{warner_Janus}, the situation is more interesting when the turning points lie closer to the boundaries of different regions. The space of solutions in this case is shown in figure \ref{N1_G2_space1} on the right panel. As the turning point moves from yellow region to grey region, the solution changes from SYM/SYM Janus to SYM/singularity Janus. As the turning point lies closer to the boundary between grey and red regoins, the SYM/singularity Janus approach the $G_2$ critical point. When the turning point lies on this boundary, the solution interpolates between SYM phase, $G_2$ critical point and a singularity. 
\\
\indent On the other hand, solutions with turning points on the red region interpolate between singularities and approach the $G_2$ critical point as the turning points come close to the boundary between red and grey regions. Examples of these solutions are shown by the green line in figure \ref{G2_sol_gen1}. In the figure, we have also given SYM/SYM Janus solutions shown by black and blue lines with the turning points in yellow region and on the boundary between yellow and grey regions, respectively. It should be noted that although the turning point of the solution shown by the blue line is on the boundary between yellow and grey regions, the solution is not singular and does not approach the $G_2$ critical point.   

\begin{figure}
\centering
    \includegraphics[width=0.55\linewidth]{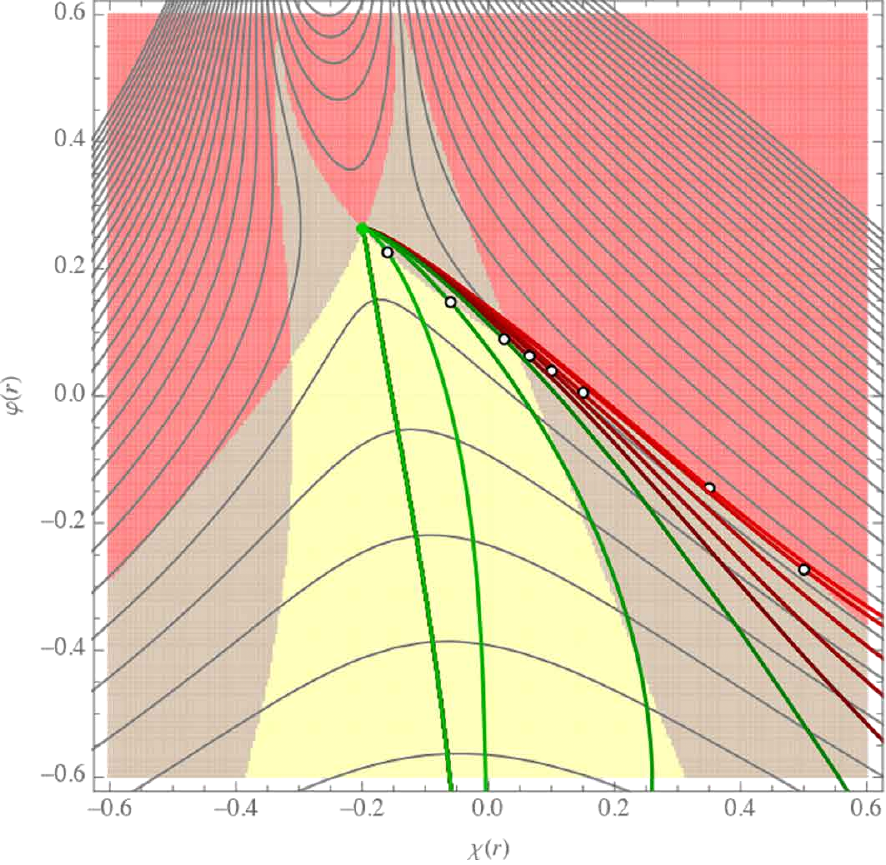}
  \caption{Space of solutions for $G_2$ symmetric Janus solutions with the turning points on the boundaries between yellow and grey regions and between red and grey regions.} 
  \label{N1_G2_space2}
\end{figure}

\begin{figure}
    \includegraphics[width=1.0\linewidth]{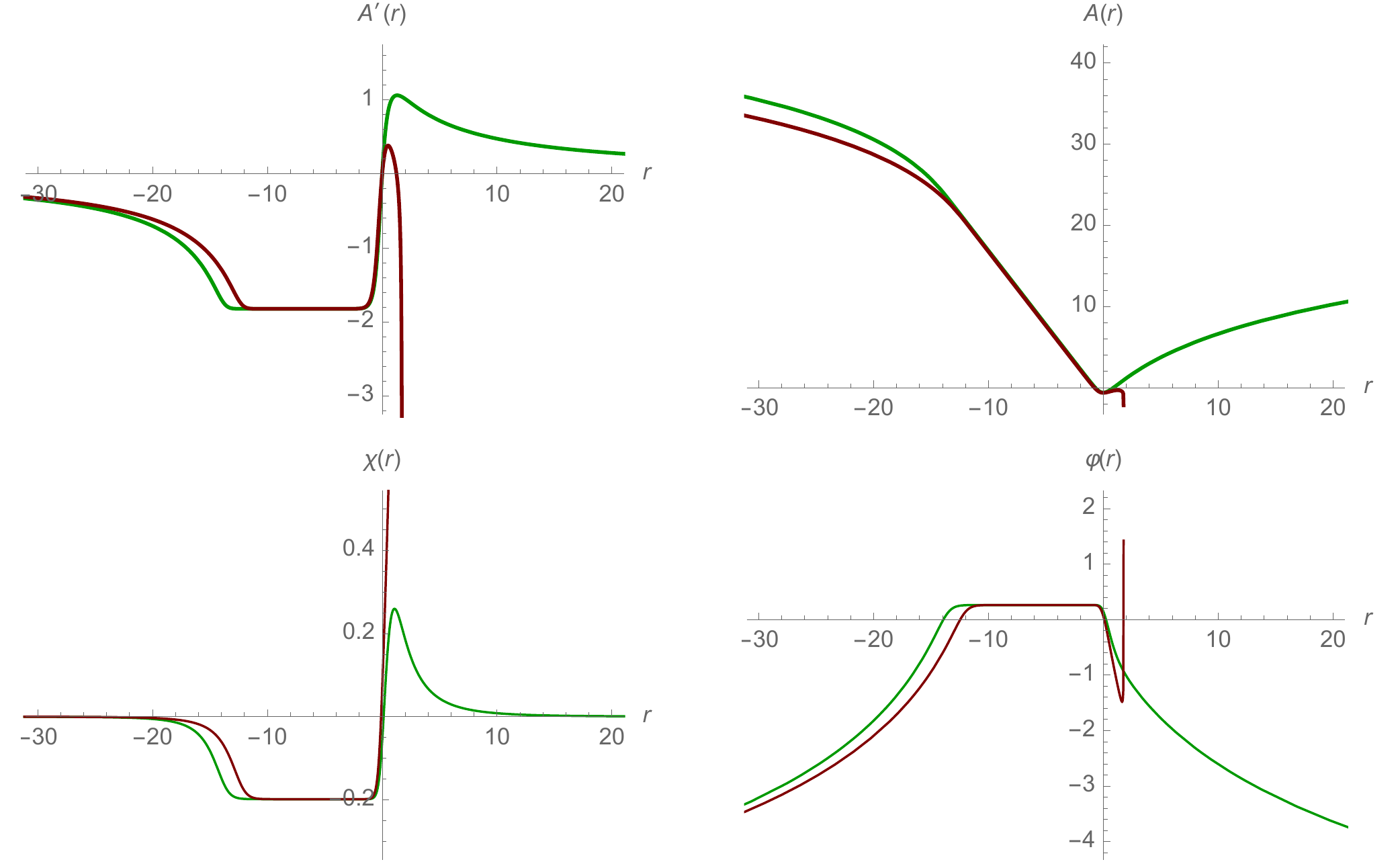}
  \caption{Examples of $N=1$ SYM/$G_2$ (green) and $G_2$/singularity (dark red) Janus solutions with $G_2$ symmetry.} 
  \label{G2_sol}
\end{figure}

Regular SYM/SYM Janus solutions that approach the $G_2$ critical point can be found if the turning points lie close to the $G_2$ critical point on the boundary between yellow and grey regions as shown in figure \ref{N1_G2_space2}. An example of these solutions is shown by the green line in figure \ref{G2_sol}. On the right side, the solution is asymptotic to the SYM phase and connects to the $G_2$ conformal phase on the left side. The $G_2$ phase then undergoes an RG flow to the SYM phase. This solution should then describe an interface between SYM and $G_2$ phases. 

\subsection{$N=1$ supersymmetric Janus solutions with $SU(3)$ symmetry}
\begin{figure}
    \includegraphics[width=1.0\linewidth]{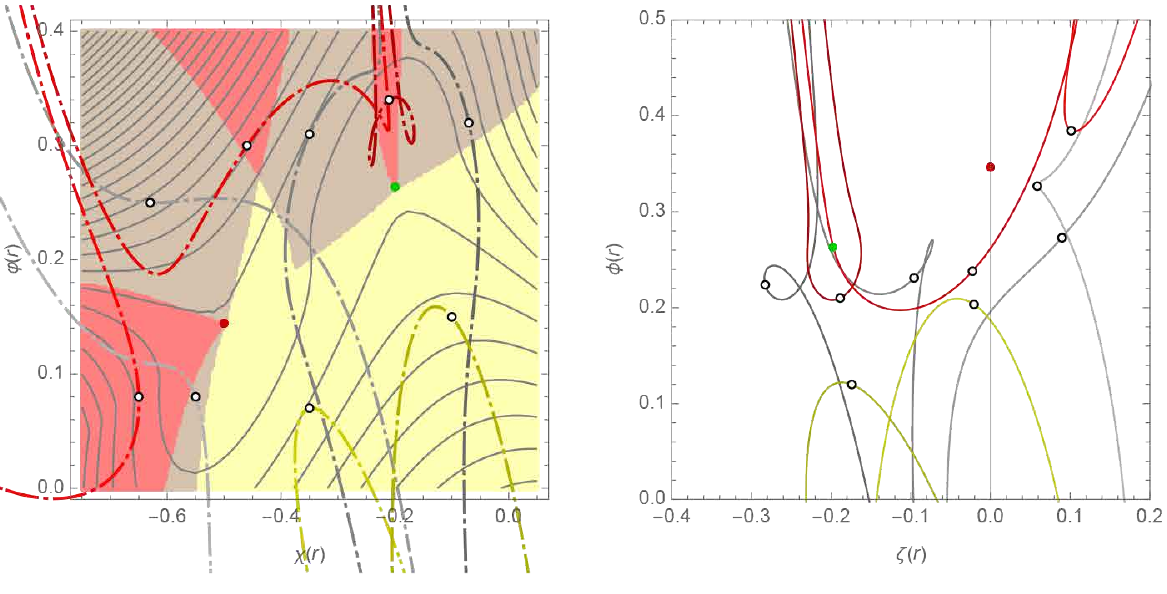}
  \caption{Space of generic $N=1$ Janus solutions with $SU(3)$ symmetry. Red and green dots represent $U(3)$ and $G_2$ critical points, respectively.} 
  \label{SU3_space}
\end{figure}

\begin{figure}
    \includegraphics[width=1.0\linewidth]{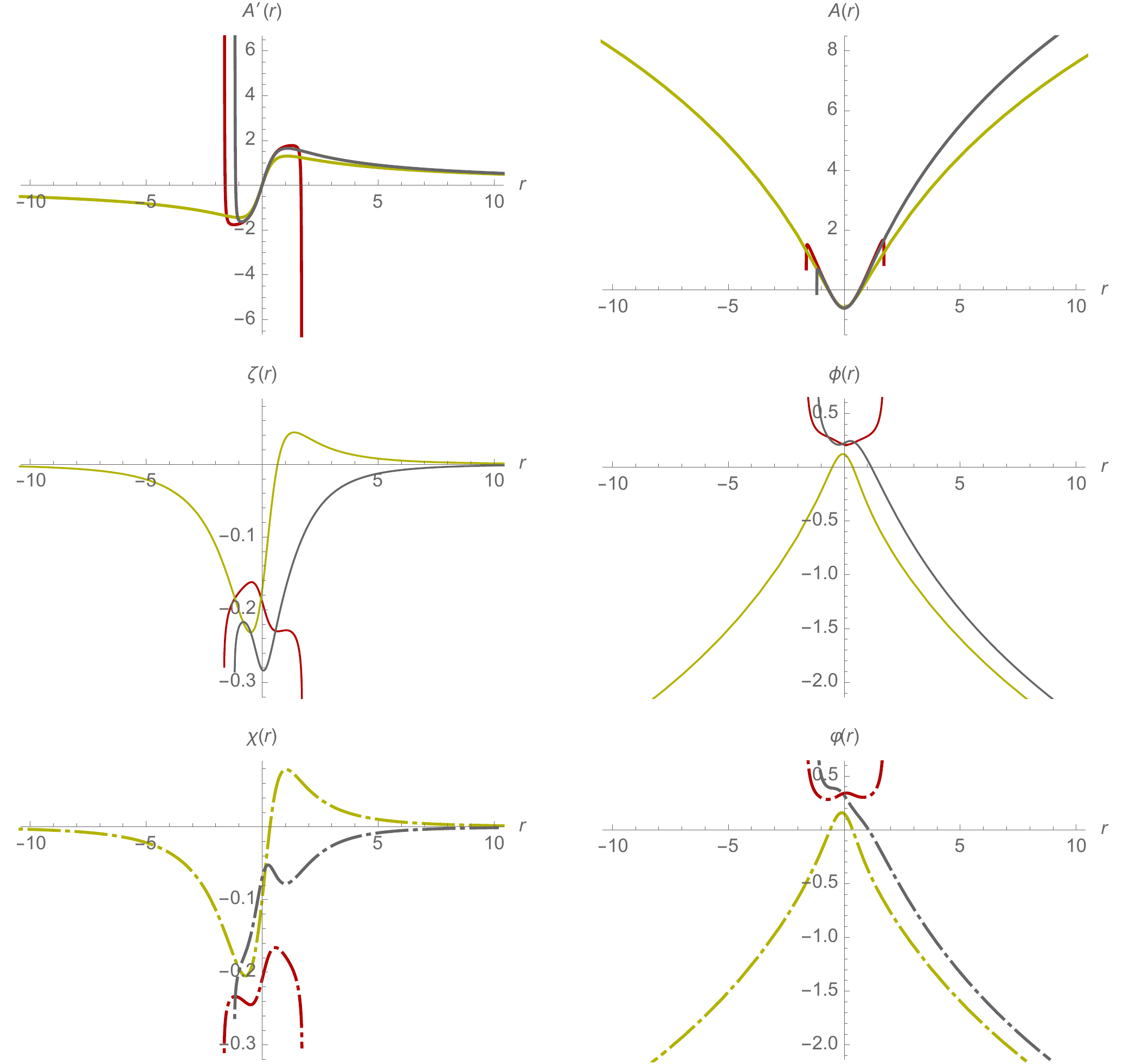}
  \caption{Examples of SYM/SYM (yellow), SYM/singularity (grey) and singularity/singularity (red) Janus solutions with $N=1$ supersymmetry and $SU(3)$ symmetry.} 
  \label{SU3_generic_sol}
\end{figure}

We now move to solutions with $SU(3)$ symmetry given by the following conditions
\begin{equation}
\begin{aligned}
& \zeta_1 = \zeta_2 \equiv \zeta \\
& \phi_1 = \phi_2 \equiv \phi \\
& \zeta_3 = \chi  \\
& \phi_3 = \varphi\, .
\end{aligned}
\end{equation}
Some examples of Janus solutions involving SYM, $U(3)$ and $SU(3)$ conformal phases have already been given in \cite{Minwoo_4DN8_Janus}, so we will mainly discuss new types of solutions.
\\
\indent In this case, there are four independent scalars $\varphi$, $\phi$, $\chi$ and $\zeta$. Various possible solutions are shown in figure \ref{SU3_space}. As can be seen from the figure, the space of solutions with turning points on different regions has a similar structure to the previous cases. There are SYM/SYM, SYM/singularity and singularity/singularity Janus solutions with turning points on yellow, grey and red regions, respectively. Examples of these solutions are given in figure \ref{SU3_generic_sol}.

\begin{figure}
    \includegraphics[width=1.0\linewidth]{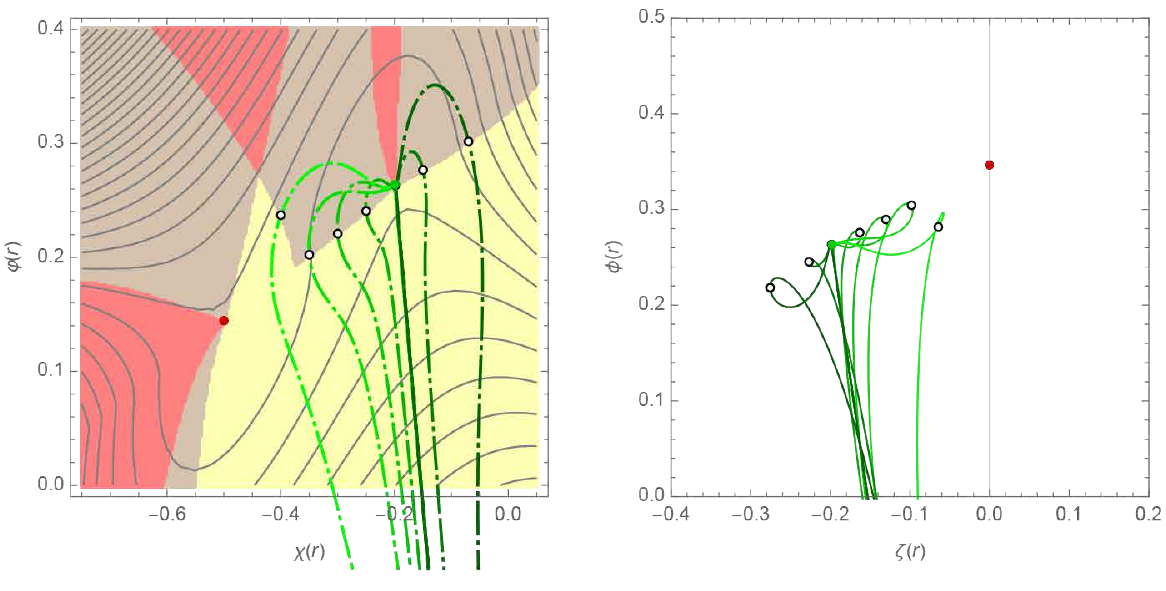}
  \caption{Space of $N=1$ Janus solutions with $SU(3)$ symmetry that approach the $G_2$ critical point.} 
  \label{SU3_G2space}
\end{figure}

\begin{figure}
    \includegraphics[width=1.0\linewidth]{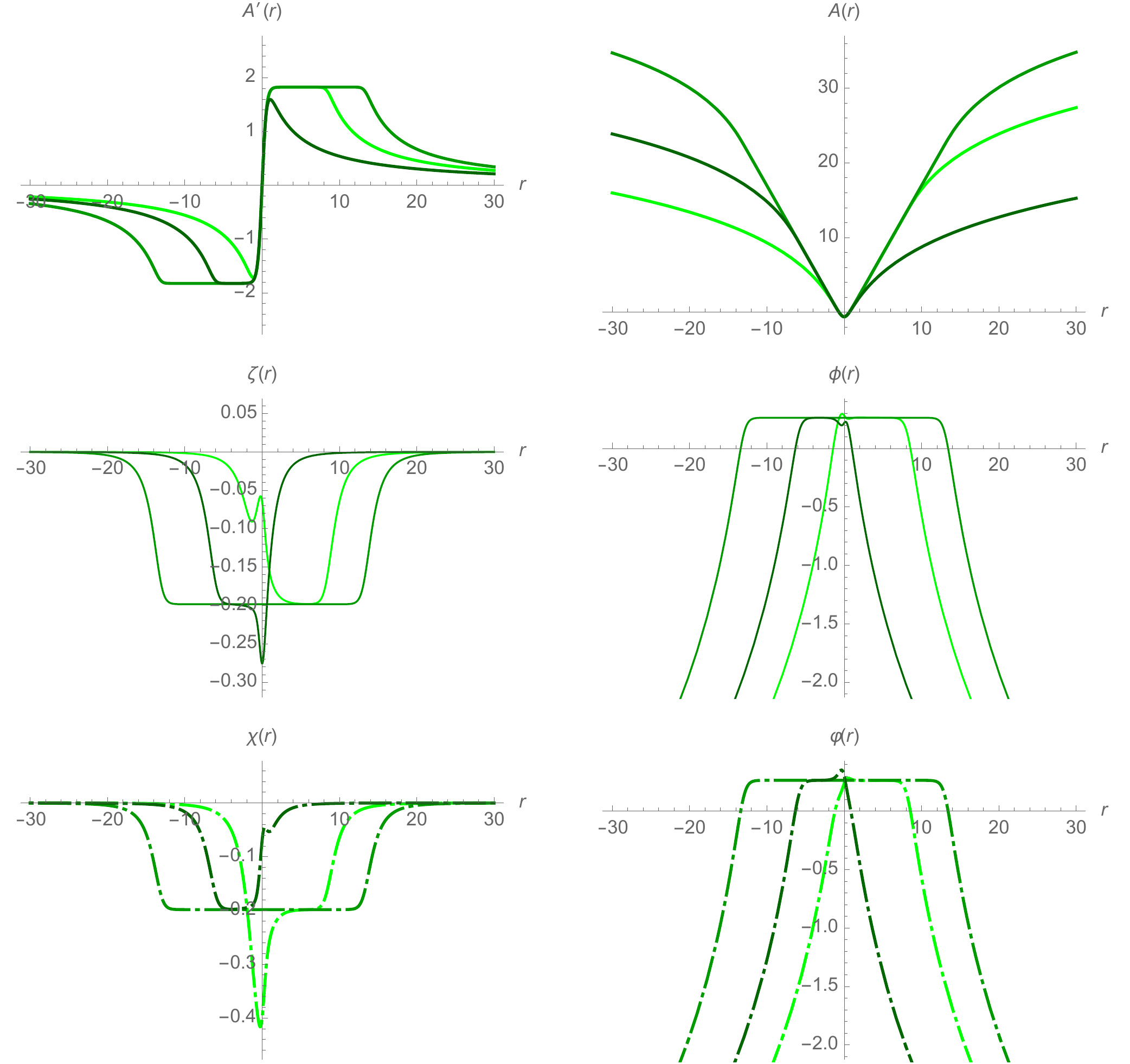}
  \caption{Examples of $G_2$/SYM (dark green), SYM/$G_2$ (light green) and $G_2$/$G_2$ (green) $N=1$ Janus solutions with $SU(3)$ symmetry.} 
  \label{SU3_G2_sol}
\end{figure}

In addition, there are solutions that involve two supersymmetric $AdS_4$ critical points with $U(3)$ and $G_2$ symmetries. If the turning points lie on the boundary between yellow and grey regions, the solutions approach the $G_2$ critical point as shown in figures \ref{SU3_G2space} and \ref{SU3_G2_sol}. There are solutions in which the SYM phase undergoes an RG flow to the $G_2$ conformal phase on one side (dark and light green lines) or on both side (green line) corresponding respectively to SYM/$G_2$ and $G_2$/SYM or $G_2$/$G_2$ Janus configurations. These solutions should describe various conformal interfaces within $N=8$ SYM on D2-branes.   

\begin{figure}
    \includegraphics[width=1.0\linewidth]{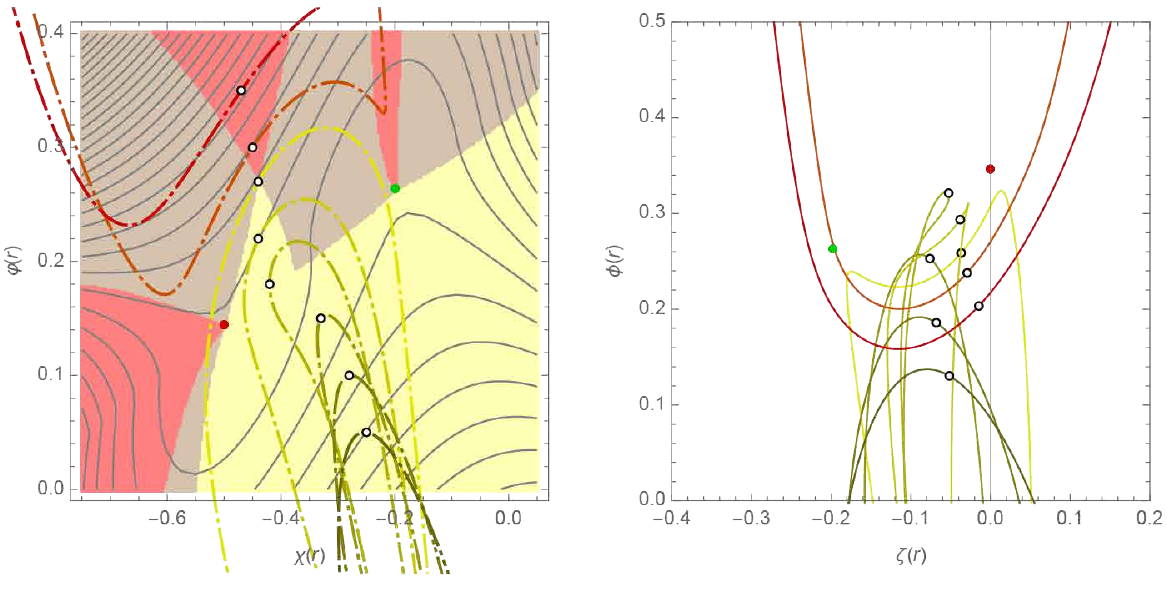}
  \caption{Space of $N=1$ Janus solutions with $SU(3)$ symmetry that approach $U(3)$ and $G_2$ critical points.} 
  \label{U3_G2space}
\end{figure}

\begin{figure}
    \includegraphics[width=1.0\linewidth]{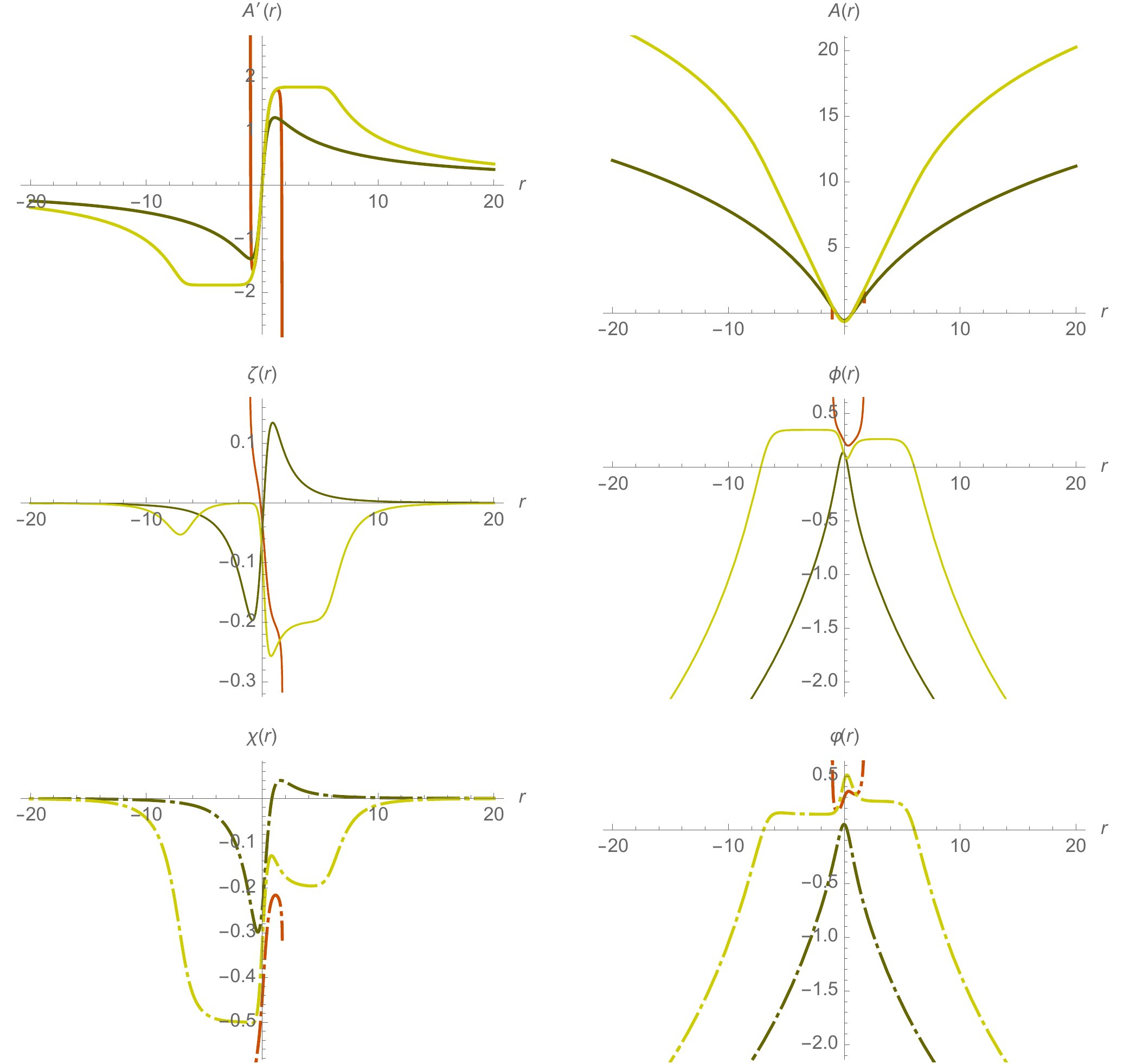}
  \caption{$G_2/U(3)$ (yellow), SYM/SYM (dark yellow) and singular (dark red) $N=1$ Janus solutions with $SU(3)$ symmetry.} 
  \label{U3_G2_sol}
\end{figure}

When the turning point lies on the boundary between yellow, grey and red regions, the solution interpolates between SYM phases and approaches both $G_2$ and $U(3)$ critical points as shown in figures \ref{U3_G2space} and \ref{U3_G2_sol}. This solution is given by the yellow line in figure \ref{U3_G2_sol}. Since the radii of $G_2$ and $U(3)$ critical points are slightly different for the chosen numerical values of the relevant parameters, the critical points on the two side of the turning point in the radial profile of $A'$ appear to be the same. However, the radial profiles of scalars clearly show that the critical points on both side of the interface are indeed different. Accordingly, this solution should describe $G_2/U(3)$ interface. For comparison, we have also given generic SYM/SYM (dark yellow) and singular (dark red) Janus solutions in figure \ref{U3_G2_sol}.  

\begin{figure}
    \includegraphics[width=1.0\linewidth]{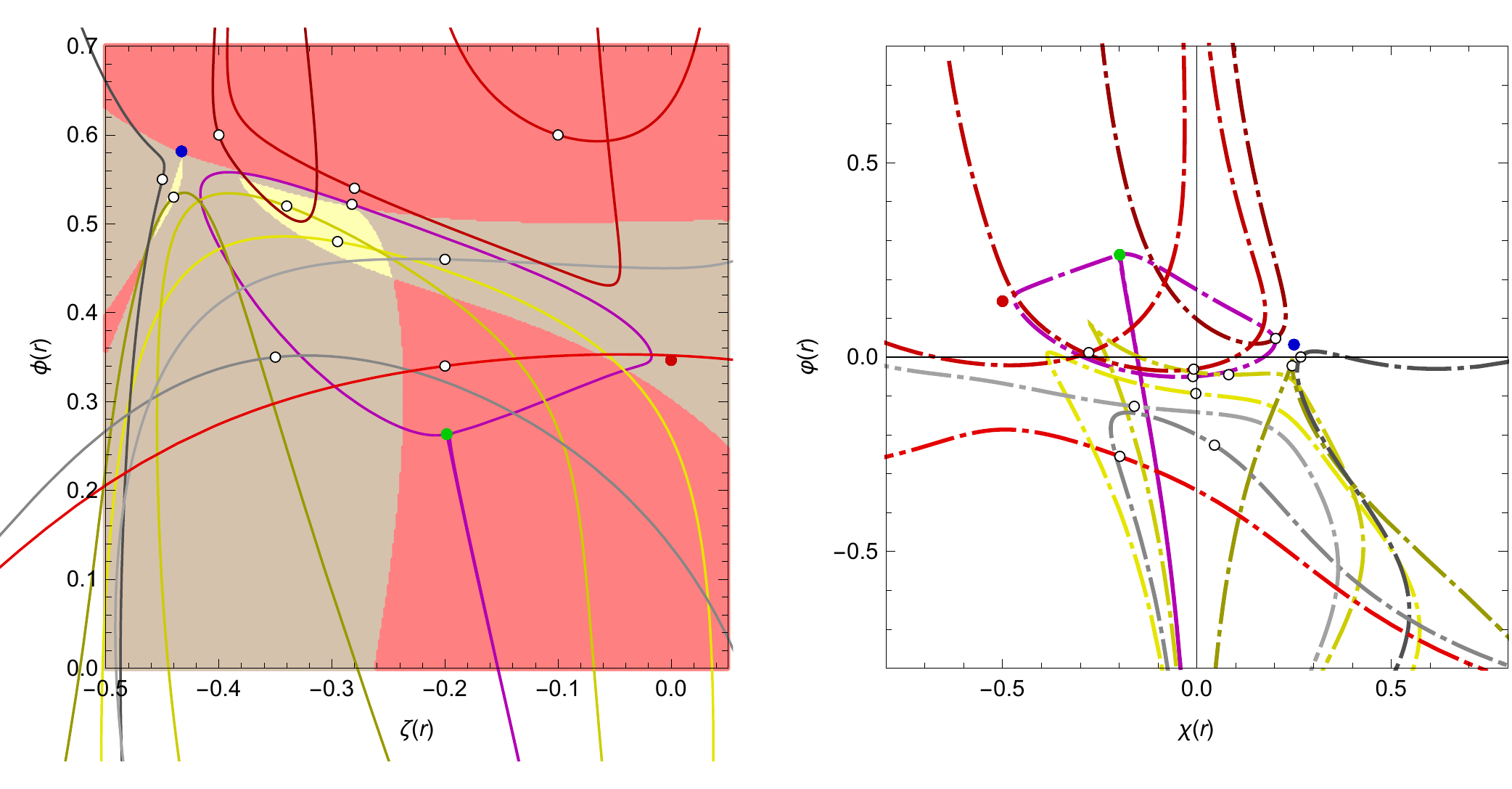}
  \caption{Space of $N=1$ Janus solutions with $SU(3)$ symmetry that approach $U(3)$ and $SU(3)$ critical points.} 
  \label{SU3_U3space}
\end{figure}

\begin{figure}
    \includegraphics[width=1.0\linewidth]{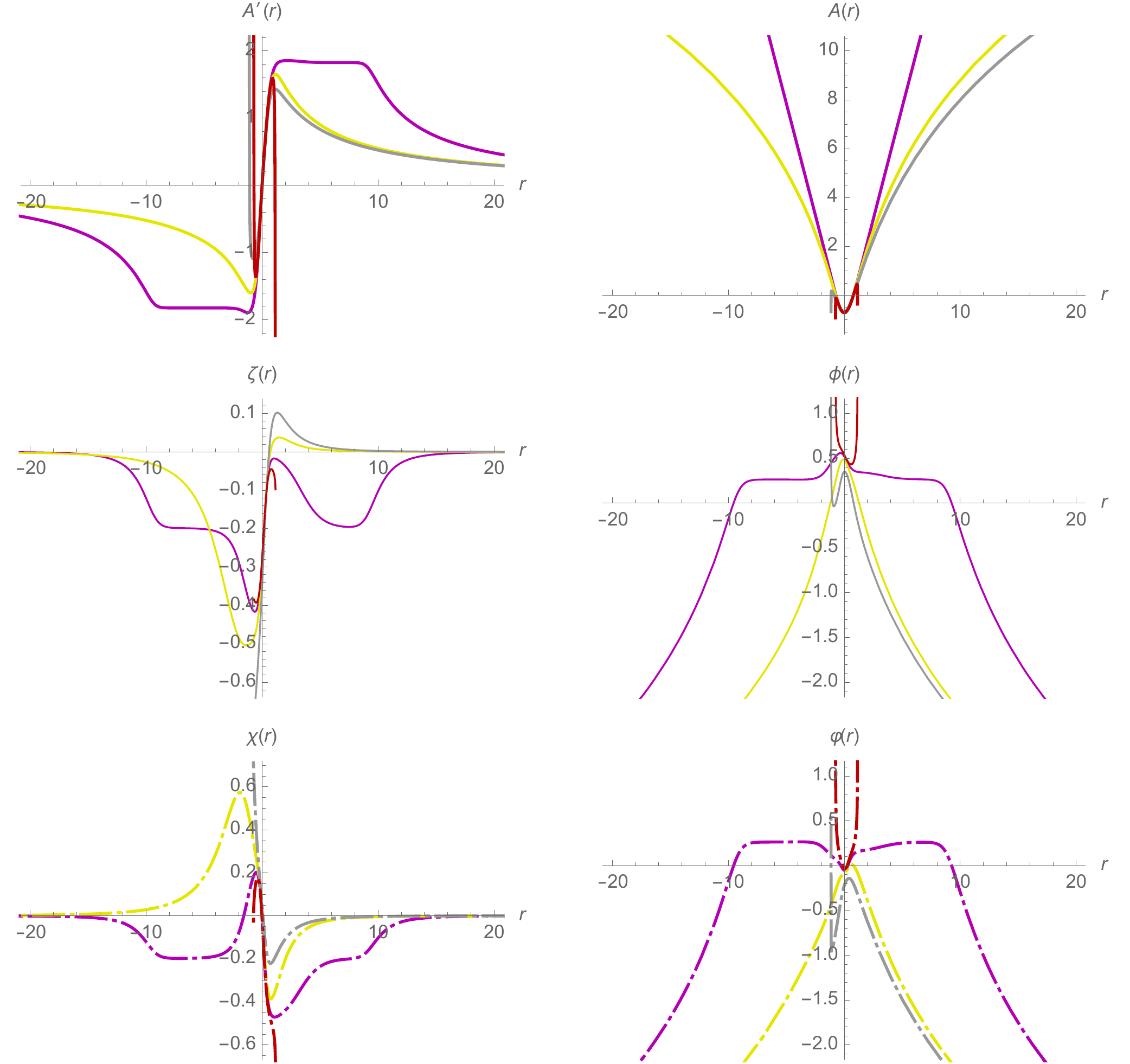}
  \caption{$U(3)/SU(3)$ (purple), $G_2/U(3)$ (yellow), SYM/singularity (grey) and singular (red) $N=1$ Janus solutions with $SU(3)$ symmetry.} 
  \label{SU3_U3_sol}
\end{figure}

Finally, there are also solutions that approach $SU(3)$ critical point as shown in figure \ref{SU3_U3space} with green, red and blue dots representing $G_2$, $U(3)$ and $SU(3)$ critical points, respectively. Examples of various solutions are given in figure \ref{SU3_U3_sol}. Among these solutions, there is an interesting solution shown by the purple line in figure \ref{SU3_U3_sol} with the turning point on the boundary between yellow, grey and red regions. The solution interpolates between SYM and $G_2$ phases and approaches arbitrarily close to $SU(3)$ and $U(3)$ critical points. The turning point lies between the $SU(3)$ and $U(3)$ phases. On one side of the interface, the SYM phase undergoes an RG flow to $G_2$ and $SU(3)$ conformal phases while on the other side, the SYM phase flows to $G_2$ and $U(3)$ phases. Therefore, we expect this solution to describe a conformal interface between $SU(3)$ and $U(3)$ phases.   

\subsection{$N=1$ supersymmetric Janus solutions with $SO(3)$ symmetry}
\begin{figure}
    \includegraphics[width=1.0\linewidth]{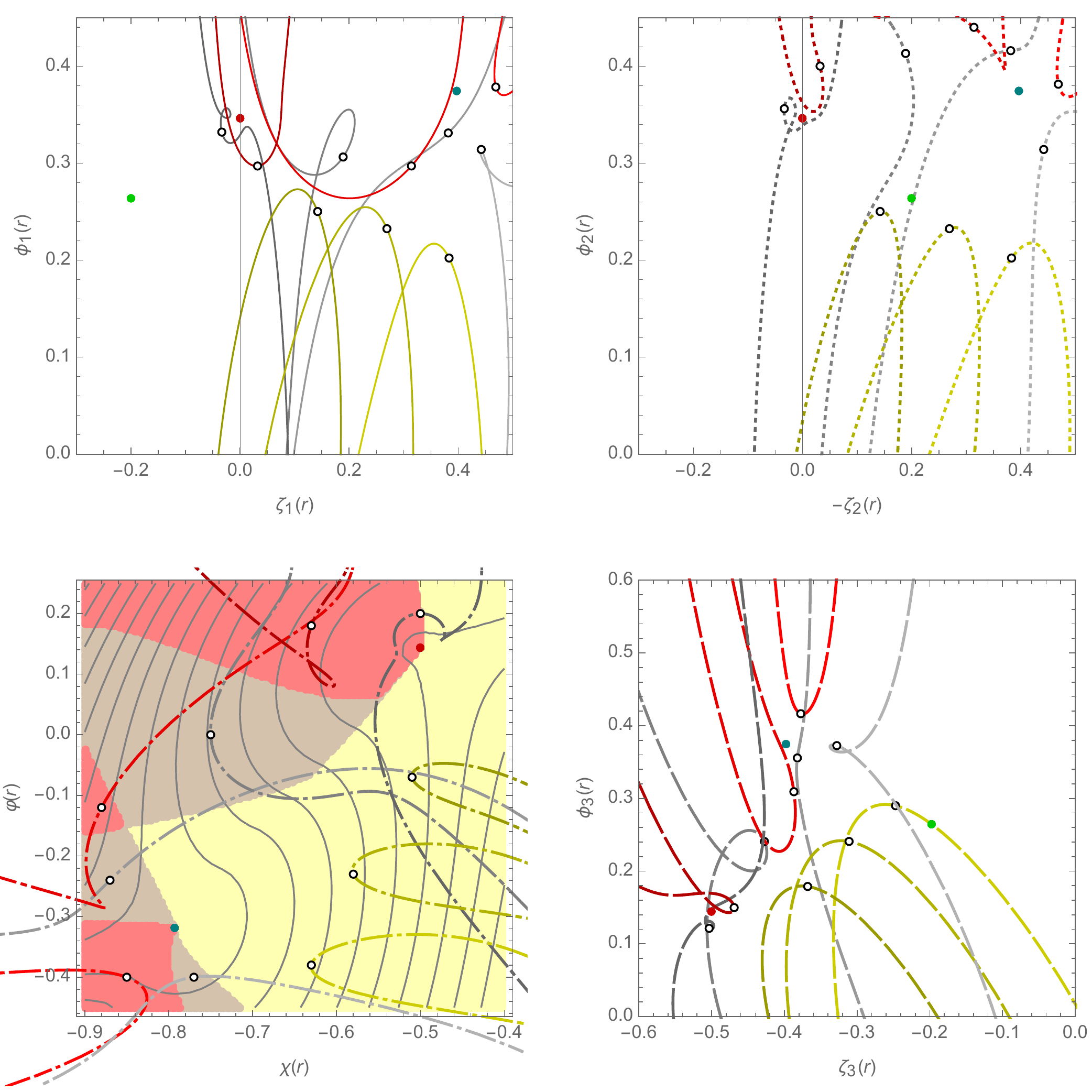}
  \caption{Space of generic $N=1$ Janus solutions with $SO(3)$ symmetry. Red and dark green dots represent $U(3)$ and $SO(4)$ critical points while green dot corresponds to $G_2$ critical point.} 
  \label{SO3_space}
\end{figure}

\begin{figure}
  \centering
  \begin{subfigure}[b]{1.0\linewidth}
    \includegraphics[width=\linewidth]{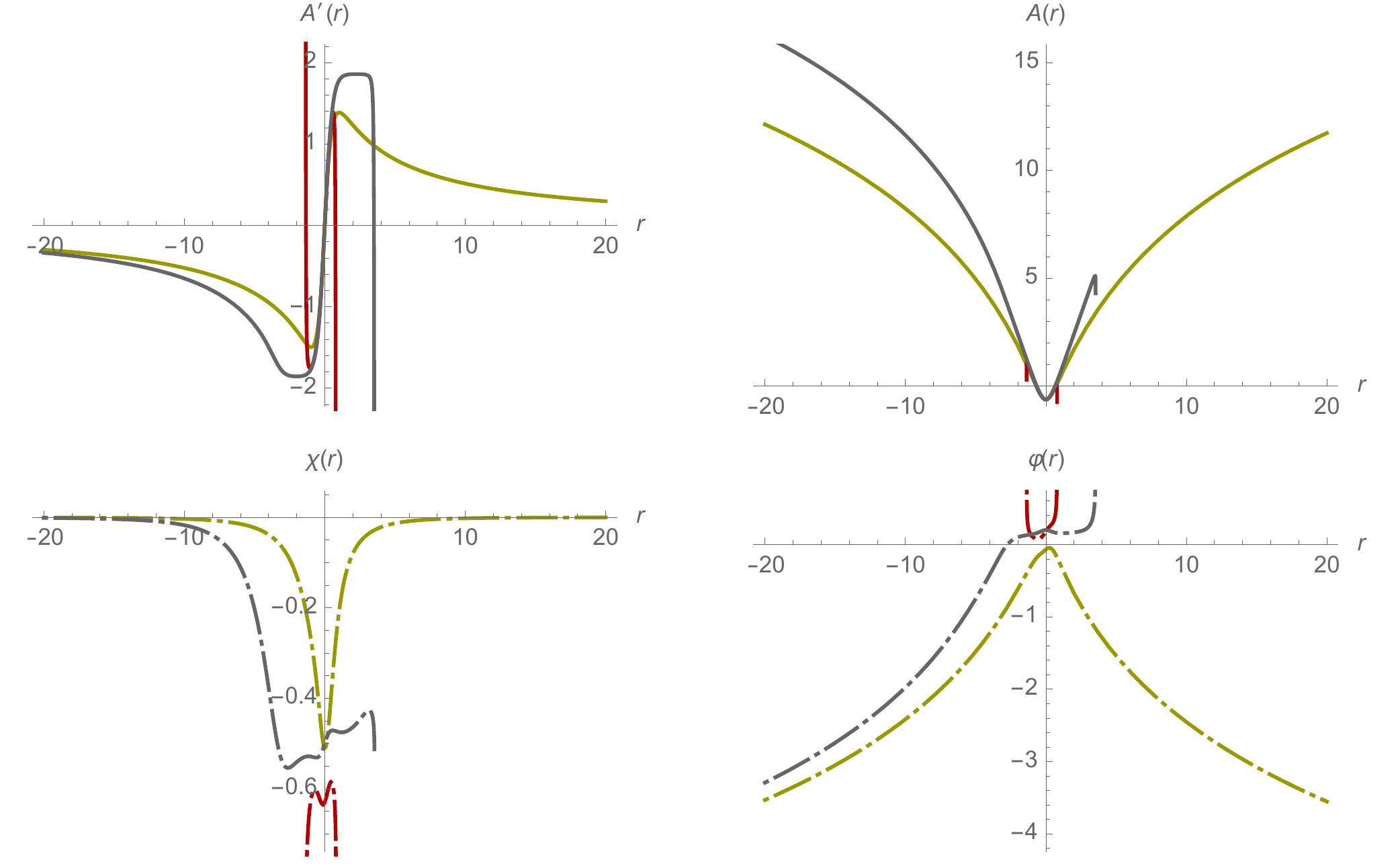}
  \end{subfigure}\\
  \begin{subfigure}[b]{1.0\linewidth}
    \includegraphics[width=\linewidth]{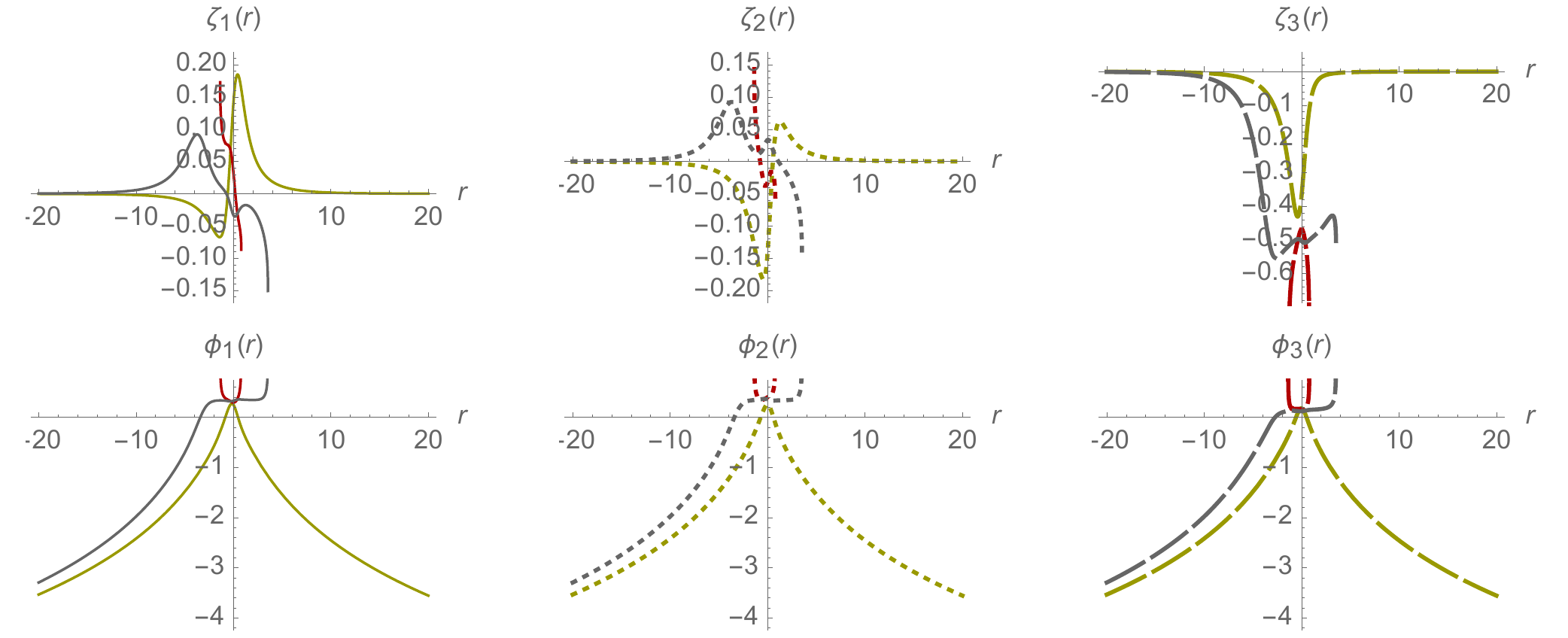}
  \end{subfigure}
\caption{SYM/SYM (yellow), SYM/singularity (grey) and singular (dark red) $N=1$ Janus solutions with $SO(3)$ symmetry.} 
  \label{SO3_generic_sol}
\end{figure}

We finally consider Janus solutions in the full $SO(3)$ invariant scalar sector parametrized by all eight scalars. With more complicated BPS equations and some difficulty in numerical analysis, the same procedure as in all the previous sections gives the space of generic solutions as in figure \ref{SO3_space}. The solutions are similar to other cases namely SYM/SYM, SYM/singularity and singular Janus solutions for turning points on yellow, grey and red regions, respectively. Examples of these solutions are shown in figure \ref{SO3_generic_sol}. 

\begin{figure}
    \includegraphics[width=1.0\linewidth]{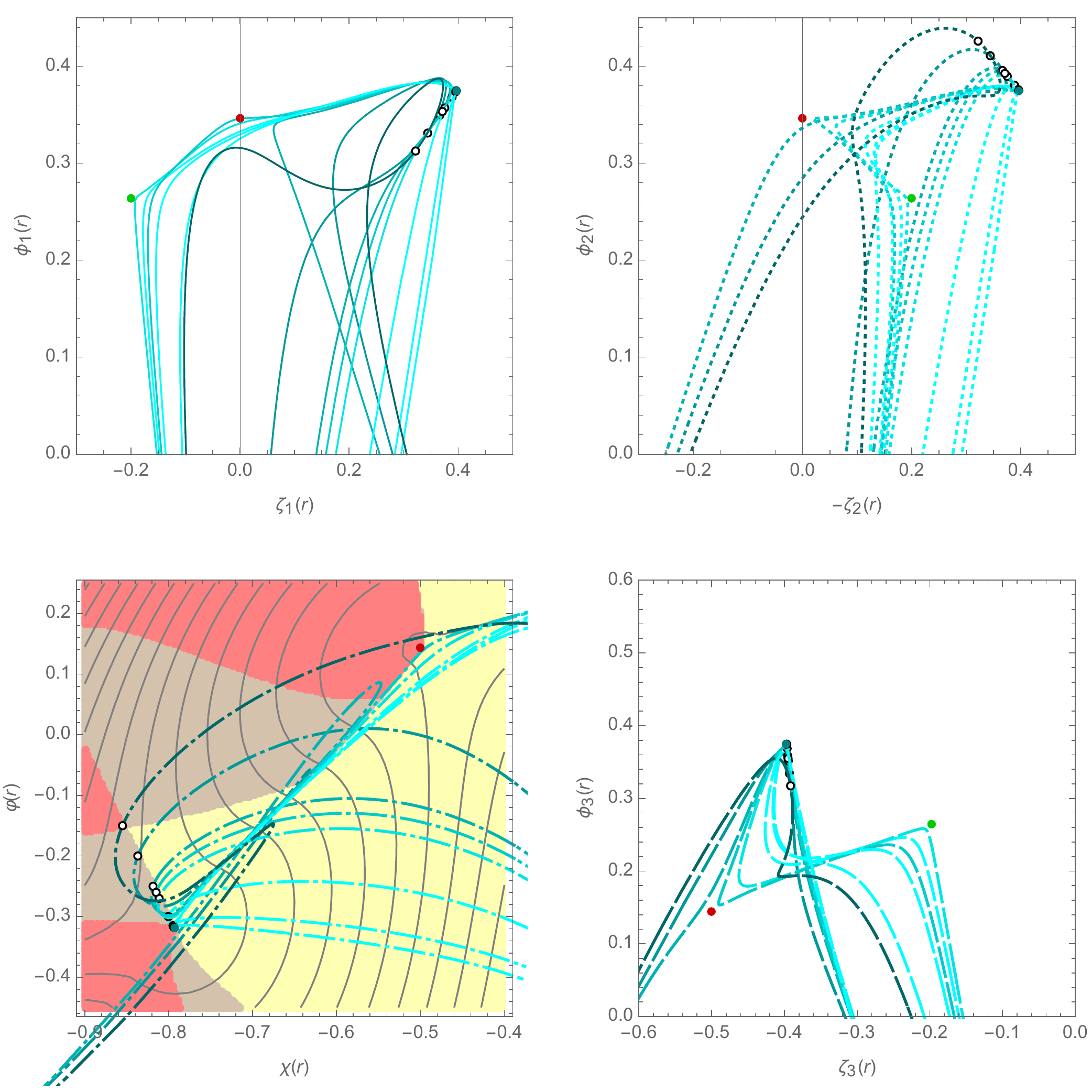}
  \caption{Space of $N=1$ Janus solutions with $SO(3)$ symmetry and turning points on the boundary between yellow and grey regions toward the $SO(4)$ critical point.} 
  \label{SO3_SO4space}
\end{figure}

\begin{figure}
  \centering
  \begin{subfigure}[b]{1.0\linewidth}
    \includegraphics[width=\linewidth]{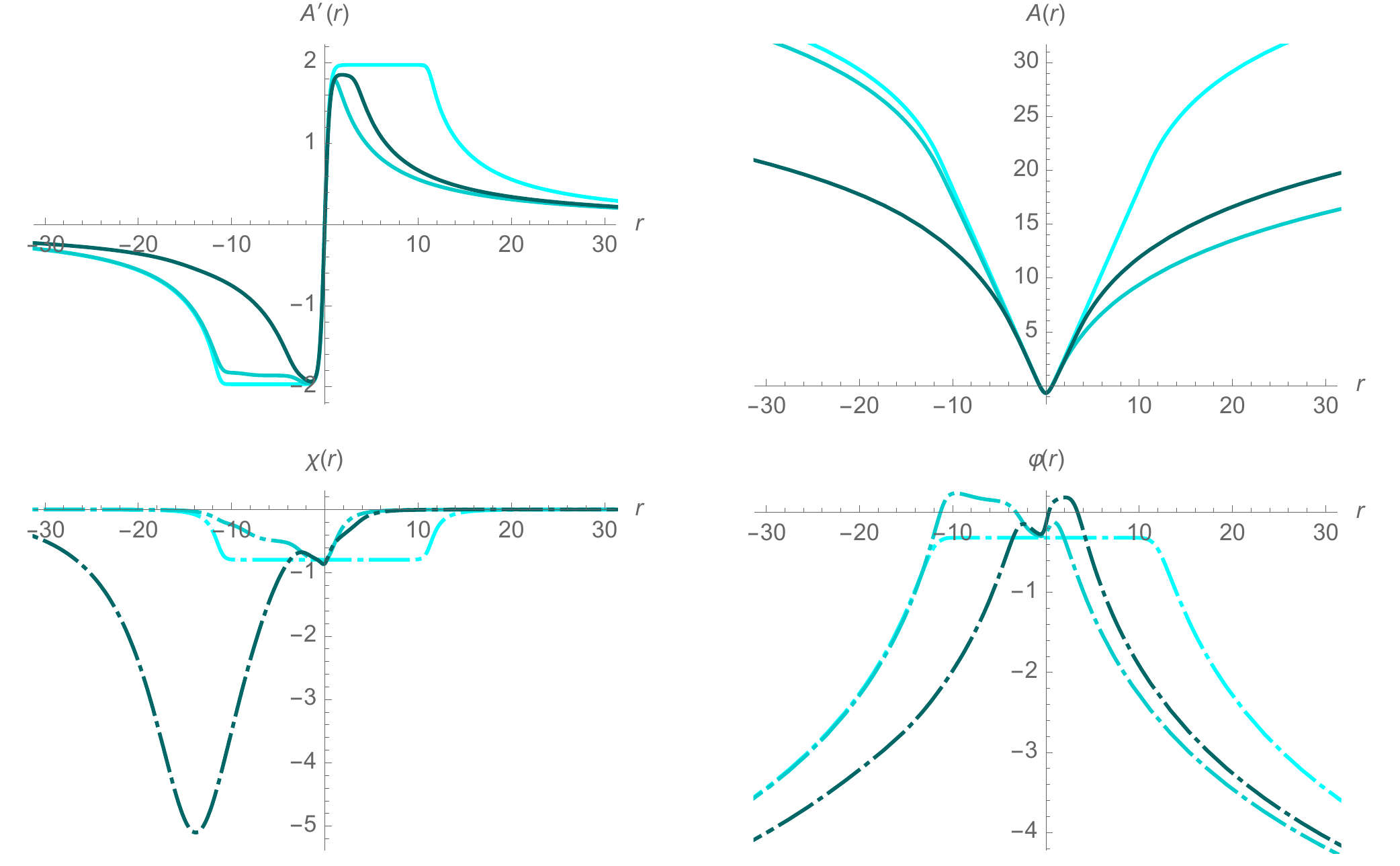}
  \end{subfigure}\\
  \begin{subfigure}[b]{1.0\linewidth}
    \includegraphics[width=\linewidth]{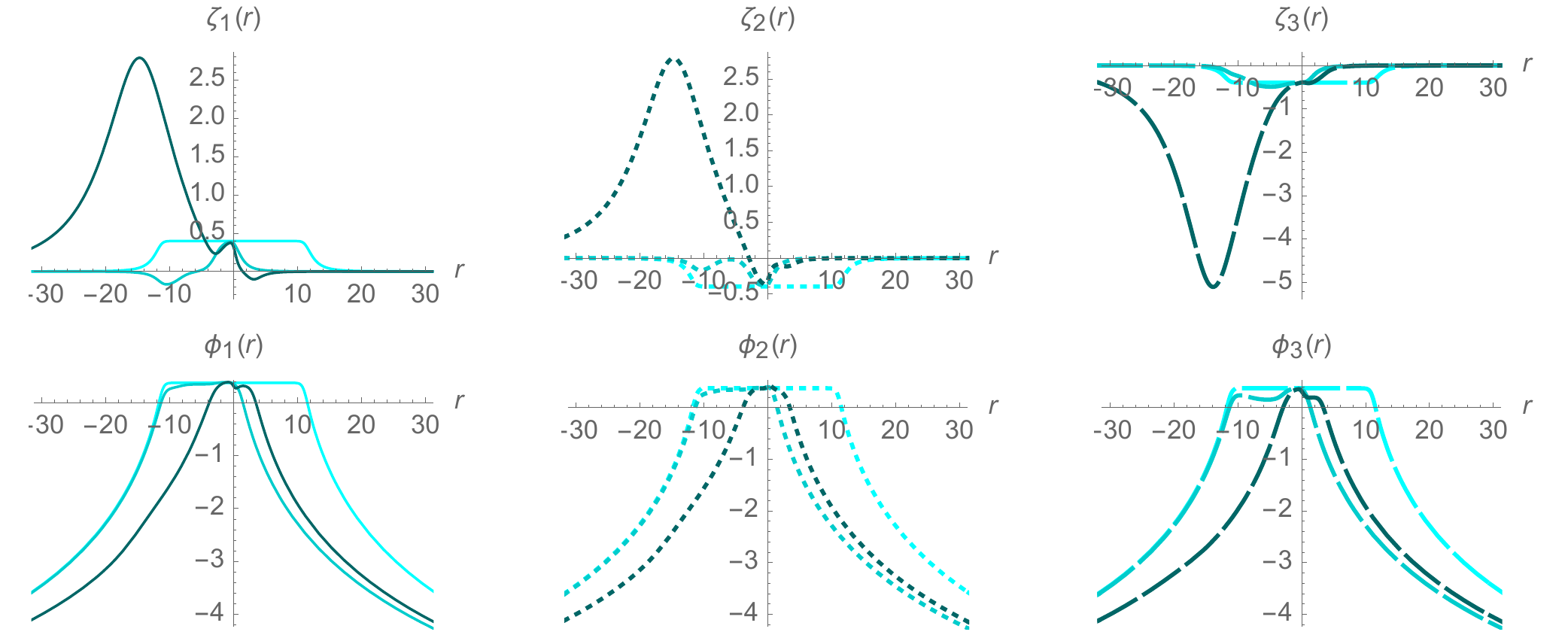}
  \end{subfigure}
 \caption{SYM/SYM (dark blue), $SO(4)/SO(4)$ (light blue) and SYM/$SO(4)$ (blue) $N=1$ Janus solutions with $SO(3)$ symmetry.} 
  \label{SO3_SO4_sol}
\end{figure}

We now move to more interesting solutions that involve one or more conformal fixed points. When the turning points lie on the boundary between yellow and grey regions toward the $SO(4)$ critical point, the solutions approach $SO(4)$ critical point as shown in figure \ref{SO3_SO4space}. An example of this type of solutions is shown by the dark blue line in figure \ref{SO3_SO4_sol}. As the turning points move closer to the $SO(4)$ critical point, the solutions will describe $SO(4)/SO(4)$ Janus solutions with the SYM phases on both sides undergoing RG flows to the $SO(4)$ conformal phases as shown by the light blue line in figure \ref{SO3_SO4_sol}. The blue line in figure \ref{SO3_SO4_sol} corresponds to a solution with SYM phase on the right and SYM/$U(3)/SO(4)$ phases on the left. In this solution, the $SO(4)$ phase on the left of the interface flows to the $U(3)$ and SYM phases. We then expect this solution to describe a SYM/$SO(4)$ interface. 

\begin{figure}
    \includegraphics[width=1.0\linewidth]{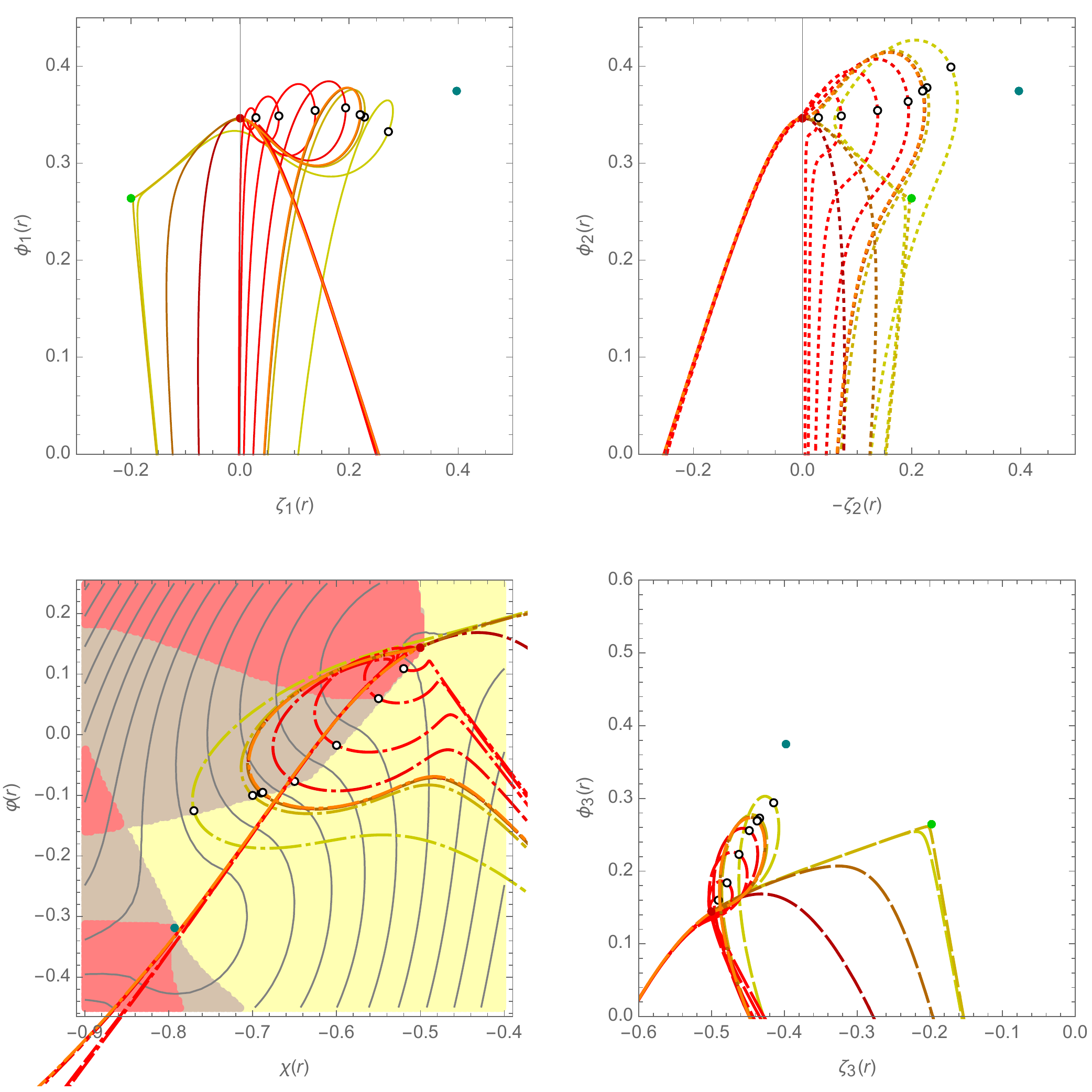}
  \caption{Space of $N=1$ Janus solutions with $SO(3)$ symmetry and turning points on the boundary between yellow and grey regions toward the $U(3)$ critical point.} 
  \label{SO3_U3space}
\end{figure}

\begin{figure}
  \centering
  \begin{subfigure}[b]{1.0\linewidth}
    \includegraphics[width=\linewidth]{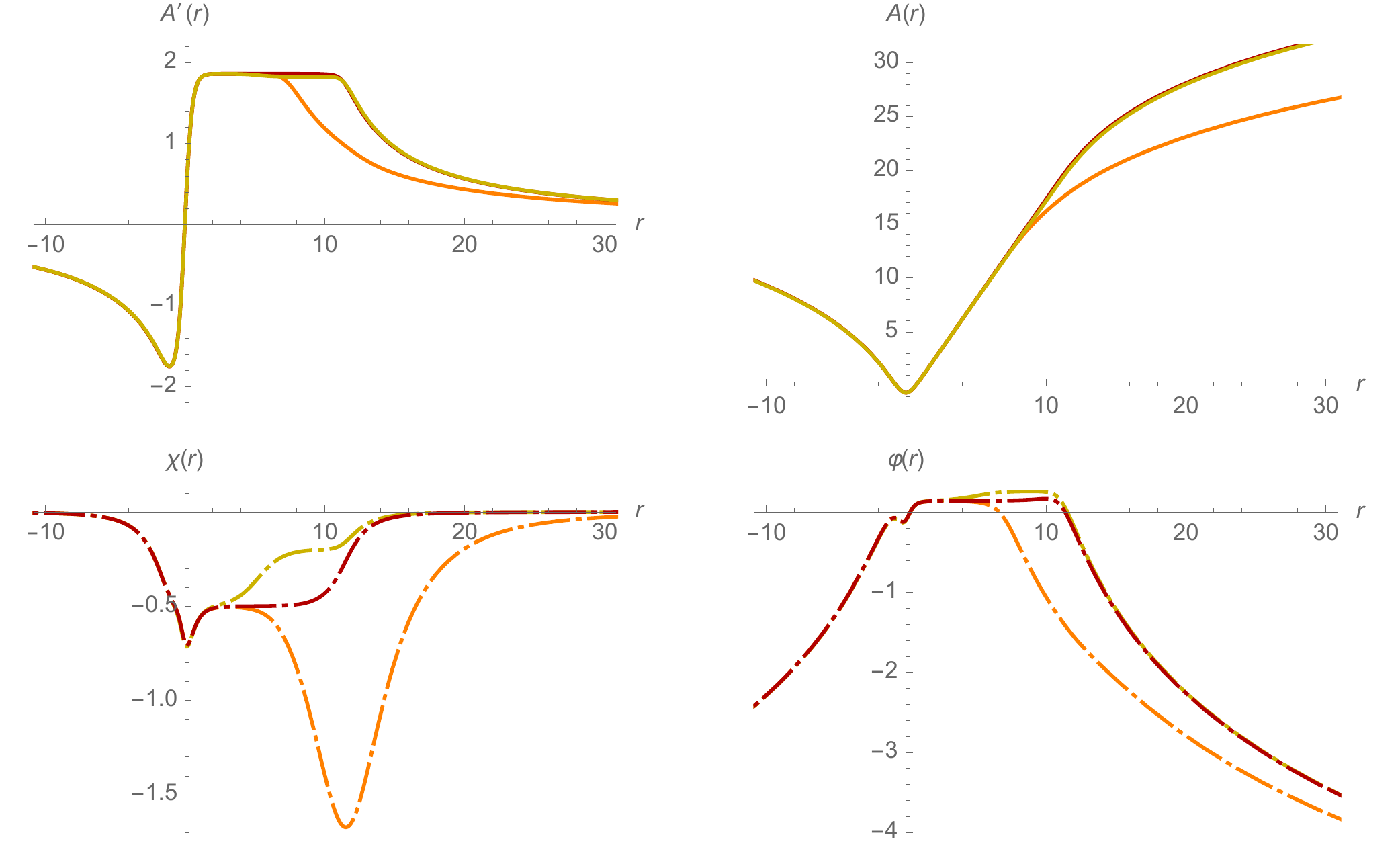}
  \end{subfigure}\\
  \begin{subfigure}[b]{1.0\linewidth}
    \includegraphics[width=\linewidth]{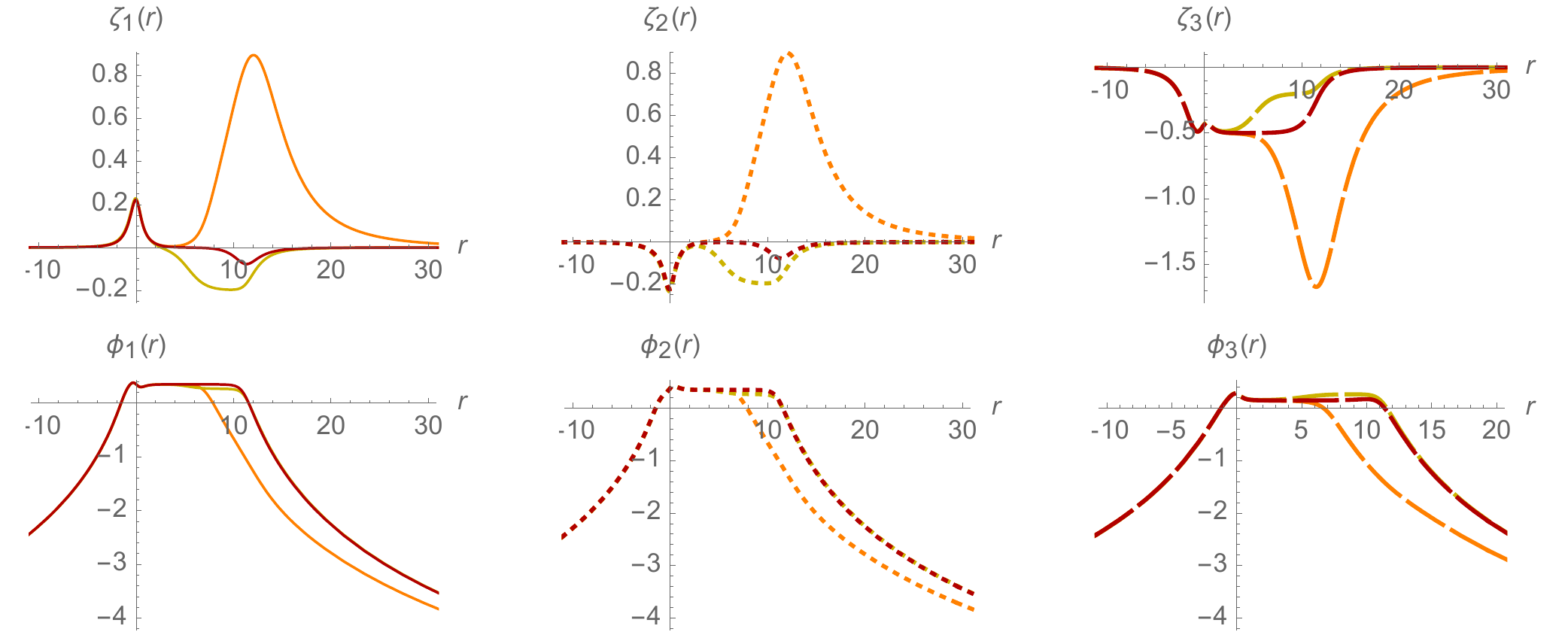}
  \end{subfigure}
 \caption{SYM/$U(3)$ $N=1$ Janus solutions with $SO(3)$ symmetry.} 
  \label{SO3_U3_sol}
\end{figure}

Similarly, when the turning points lie on the boundary between yellow and grey regions toward the $U(3)$ critical point, the solutions approach $U(3)$ critical point as shown in figure \ref{SO3_U3space}. These solutions describe $N=1$ SYM/$U(3)$ interfaces with some examples shown in figure \ref{SO3_U3_sol}. 

\begin{figure}
    \includegraphics[width=1.0\linewidth]{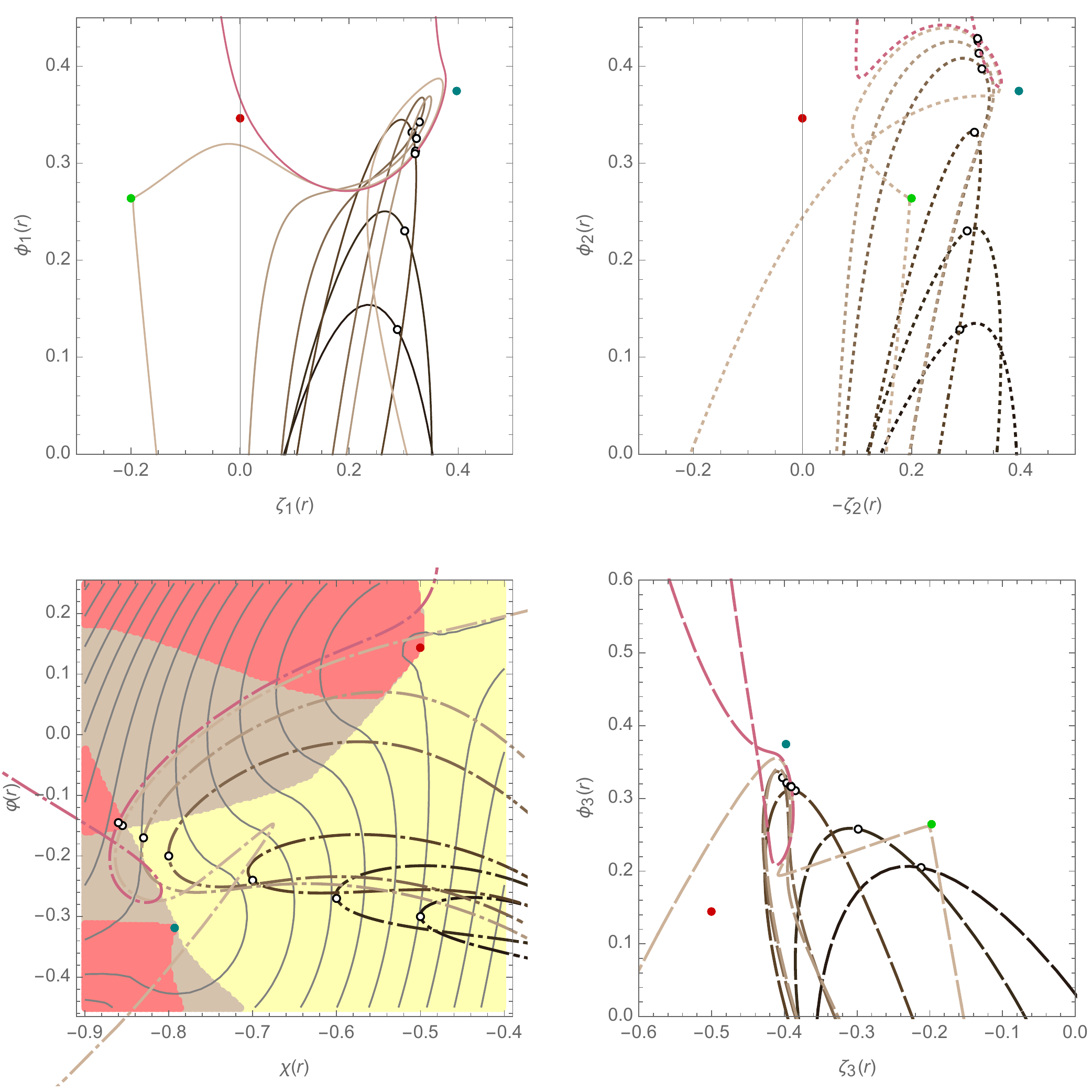}
  \caption{Space of $N=1$ Janus solutions with $SO(3)$ symmetry and turning points moving acorss the boundary between yellow, grey and red regions.} 
  \label{SO3_U3SO4space}
\end{figure}

\begin{figure}
  \centering
  \begin{subfigure}[b]{1.0\linewidth}
    \includegraphics[width=\linewidth]{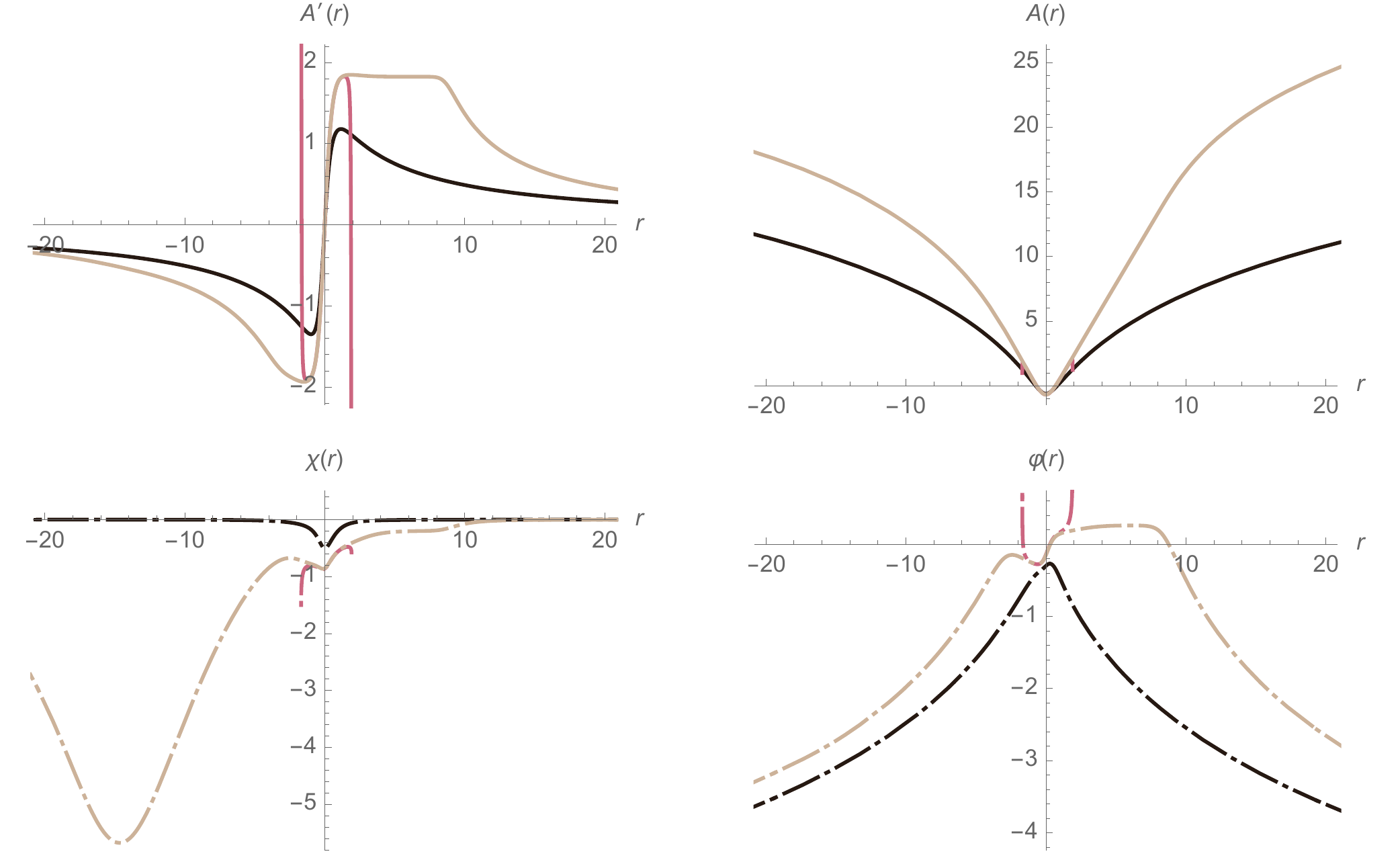}
  \end{subfigure}\\
  \begin{subfigure}[b]{1.0\linewidth}
    \includegraphics[width=\linewidth]{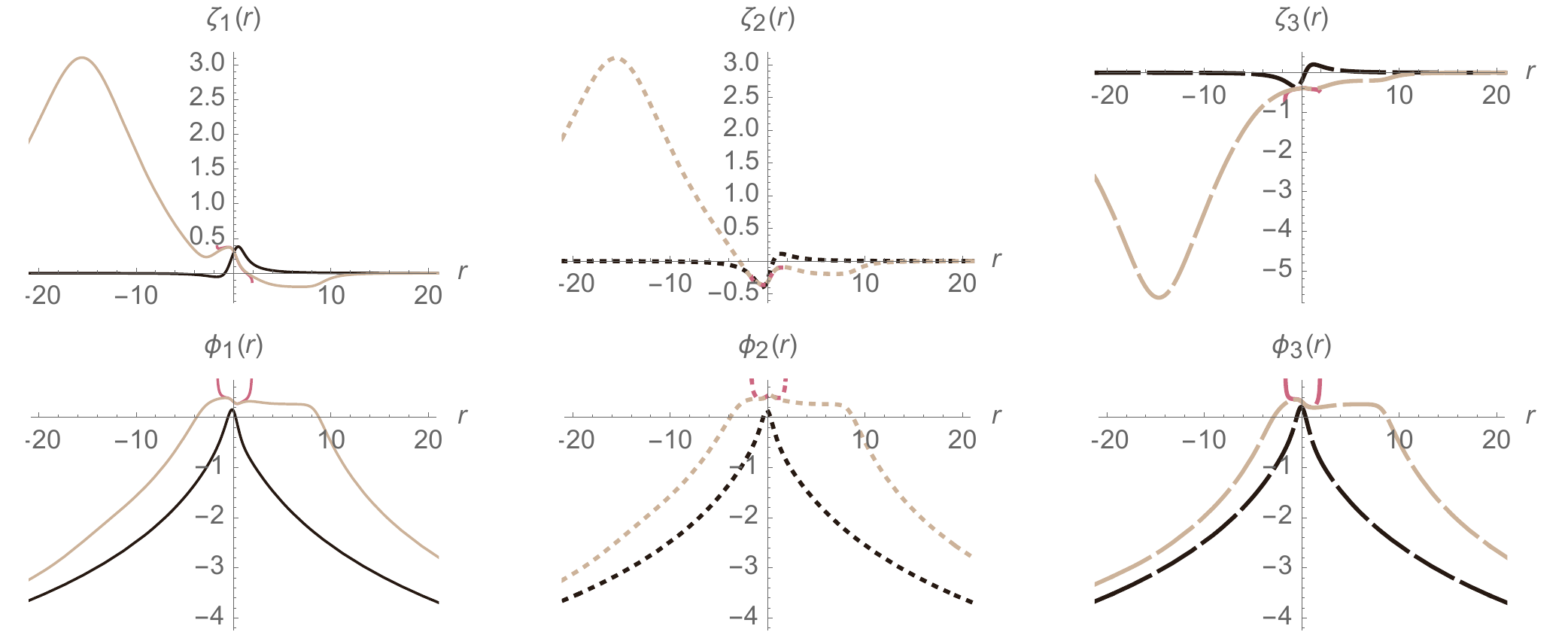}
  \end{subfigure}
 \caption{$SO(4)/U(3)$ (brown), SYM/SYM (black) and singular (red) Janus solutions with $N=1$ supersymmetry and $SO(3)$ symmetry.} 
  \label{SO3_U3SO4_sol}
\end{figure}

Finally, if the turning point lies at the boundary between yellow, red and grey regions, the solutions approach $SO(4)$ critical point on one side of the interface and the $U(3)$ phase on the other. An example of these solutions is shown by the brown line in figures \ref{SO3_U3SO4space} and \ref{SO3_U3SO4_sol}. This solution then describes an interface between $SO(4)$ and $U(3)$ conformal phases of the $N=8$ SYM. The SYM/SYM (black) and singular (red) Janus solutions are also given in figures \ref{SO3_U3SO4space} and \ref{SO3_U3SO4_sol} for comparison.   
\\
\indent We end this section by remarking that there should also be solutions approaching the $SU(3)$ $AdS_4$ critical point as in the case of $SU(3)$ symmetric solutions. However, these solutions (if exist) are very difficult to find. We have performed an intensive numerical search but have not found any examples of these solutions.

\section{Conclusions and discussions}\label{conclusion}
In this paper, we have studied supersymmetric Janus solutions of four-dimensional $N=8$ gauged supergravity with dyonic $ISO(7)$ gauge group in $SO(3)$ invariant sector. The solutions preserve $N=1,2,3$ supersymmetries and interpolate among $N=8$ SYM phase and four supersymmetric fixed points with $G_2$, $SU(3)$, $U(3)$ and $SO(4)$ symmetries. The results substantially enlarge a small number of known Janus solutions found in \cite{Minwoo_4DN8_Janus} and give rise to many new Janus solutions in four-dimensional gauged supergravities. The solutions can also be embedded in massive type IIA theory via an $S^6$ consistent truncation leading to new holographic decriptions of conformal interfaces between various phases of $N=8$ SYM theory on the world-volume of D2-branes.  
\\
\indent It would be interesting to explicitly uplift the solutions to ten dimensions and study various aspects of the conformal interfaces dual to the solutions given in this paper similar to the analysis in $SO(8)$ gauged supergravity given in \cite{Warner_N8_uplift}. Given the concrete dual field theories with $N=2,3$ supersymmetries, in terms of Chern-Simons-Matter (CSM) theories of \cite{3D_SCFT_Schwarz,3D_SCFT_Gaiotto}, in \cite{mIIA_S6_Guariano1} and a detailed map between supergravity fields and field theory operators in \cite{ISO7_N3_flow}, it is of particular interest to study the corresponding conformal interfaces in the dual field theories. Therefore, at least for the $N=2,3$ solutions, both the gravity and field theory sides are known to some extent, and matching the results from both sides would lead to more insights to the gauge/gravity duality. We hope the solutions in this paper could be useful along this line. 
\\
\indent Finally, finding other types of solutions within the dyonic $ISO(7)$ maximal gauged supergravity such as $AdS_4$ black holes and solutions describing wrapped D2-branes on two-dimensional spaces could also be interesting to investigate. Some of $AdS_4$ black hole solutions in the $N=2$ truncation have been given in \cite{Guarino_AdS2_1} and \cite{Guarino_AdS2_2}. Finding similar solutions in the $N=4$ truncation with $SO(3)$ symmetry is worth considering. This can be done along the line of the recent result \cite{N3_4_AdS4_BH} in which a number of supersymmetric $AdS_4$ black holes from $N=4$ gauged supergravity have been found. We leave these issues for future work.           
\vspace{0.5cm}\\
{\large{\textbf{Acknowledgement}}} \\
This work is supported by The Thailand Research Fund (TRF) under grant RSA6280022. We would like to thank N. Bobev for bringing reference \cite{Bobev_AdS4_ISO7} to our attention.

\end{document}